\documentclass[a4paper,11pt]{article}

\usepackage{jheppub}
\usepackage{xcolor}
\DeclareMathOperator{\li}{Li}
\DeclareMathOperator{\qli}{QLi}
\DeclareMathOperator{\Li}{Li}
\DeclareMathOperator{\QLi}{QLi}
\usepackage[dvipsnames]{xcolor}
\usepackage{tikz}
\usetikzlibrary{calc,arrows.meta,decorations.markings}
\newcommand{\crB}{\operatorname{cr}^{B}}
\usepackage{microtype}

\newcommand{\octbase}{%
  \foreach \i/\lab/\ang in {1/1/112.5, 2/2/67.5, 3/3/22.5, 4/4/-22.5,
                            5/{\bar 1}/-67.5, 6/{\bar 2}/-112.5,
                            7/{\bar 3}/-157.5, 8/{\bar 4}/157.5}{
    \coordinate (v\i) at (\ang:1);
    \node[font=\scriptsize] at (\ang:1.24) {$\lab$};
  }
  \draw[line width=0.9pt] (v1)--(v2)--(v3)--(v4)--(v5)--(v6)--(v7)--(v8)--cycle;
}
\newcommand{\dodecbase}{%
  \foreach \i/\lab in {1/1, 2/2, 3/3, 4/4, 5/5, 6/6,
                       7/{\bar 1}, 8/{\bar 2}, 9/{\bar 3},
                       10/{\bar 4}, 11/{\bar 5}, 12/{\bar 6}}{
    \pgfmathsetmacro{\ang}{105-30*(\i-1)}
    \coordinate (w\i) at (\ang:1);
    \node[font=\scriptsize] at (\ang:1.22) {$\lab$};
  }
  \draw[line width=0.8pt] (w1)--(w2)--(w3)--(w4)--(w5)--(w6)--(w7)
                          --(w8)--(w9)--(w10)--(w11)--(w12)--cycle;
}
\tikzset{
  arb/.style={postaction={decorate},
    decoration={markings,mark=at position 0.7 with {\arrow[scale=1.3]{Stealth}}},thick}
}

\title{de Sitter Wavefunction from Quadrangular Polylogarithms: Polygon Graphs}

\author[1]{Livia Ferro,}
\emailAdd{l.ferro@herts.ac.uk}

\author[1]{Tomasz {\L}ukowski,}
\emailAdd{t.lukowski@herts.ac.uk}

\author[3]{Shruti Paranjape,}
\emailAdd{shruti\_paranjape@brown.edu}

\author[2]{Lecheng Ren,}
\emailAdd{lecheng.ren@qmul.ac.uk}

\author[3,4]{Marcus Spradlin,}
\emailAdd{marcus\_spradlin@brown.edu}

\author[3]{Anastasia Volovich,}
\emailAdd{anastasia\_volovich@brown.edu}

\author[3]{He-Chen Weng}
\emailAdd{he-chen\_weng@brown.edu}

\author[1]{and Yao-Qi Zhang}
\emailAdd{y.zhang59@herts.ac.uk}

\affiliation[1]{Department of Physics, Astronomy and Mathematics,
University of Hertfordshire,
Hatfield, Hertfordshire, AL10 9AB, United Kingdom}

\affiliation[2]{Centre for Theoretical Physics, Department of Physics and Astronomy,
Queen Mary University of London,
London, E1 4NS, United Kingdom}

\affiliation[3]{Department of Physics,
Brown University,
Providence, RI 02912, USA}

\affiliation[4]{Brown Center for Theoretical Physics and Innovation,
Brown University,
Providence, RI 02912, USA}

\abstract{We present an explicit formula for the $n$-gon graph contribution to the integrand of the cosmological wavefunction for conformally coupled $\phi^3$ theory in de Sitter space. Our result relies on the fact that its symbol satisfies total compatibility with respect to the $B_{2n-1}$ cluster algebra. We introduce $B$-type quadrangular polylogarithms, constructed from centrally symmetric quadrangulations of a $4n$-gon, and show that the $n$-gon wavefunction coefficient is expressed by a natural weight-$n$ function of this class. We prove our formula by directly relating the recursive differential equation satisfied by these wavefunction coefficients to a recursive coproduct formula for $B$-type quadrangular polylogarithms, with the remaining additive constant fixed by the soft limit.}

\begin{document}
\maketitle
\flushbottom

\section{Introduction}

In recent years there has been considerable interest in the mathematical structure of cosmological observables; see for example~\cite{Arkani-Hamed:2017fdk,Benincasa:2020aoj,Melville:2021lst,De:2023xue,Benincasa:2024leu,De:2024zic,Arkani-Hamed:2024jbp,Glew:2025otn,De:2025bmf,Capuano:2025ehm,Baumann:2025qjx,Arundine:2026fbr,McLeod:2026jpz,Huang:2026tsh,Baumann:2026atn,Chowdhury:2026dwm,McLeod:2026kpo,He:2026bwn,Glazer:2026whh,Larkin:2026ayy}.
Wavefunction coefficients for a conformally coupled scalar $\phi^3$ theory in de Sitter space evaluate to generalised polylogarithm functions and hence may be studied via the tools of symbology~\cite{Goncharov:2010jf}. It has recently been observed that the symbol letters of the wavefunction coefficients associated to certain graphs are cluster variables of finite-type cluster algebras, like for certain amplitudes in $\mathcal{N}=4$ super-Yang-Mills (SYM) theory~\cite{Golden:2013xva}. Specifically, the alphabets for chain and polygon graphs consist of cluster variables of $A$- and $B$-type cluster algebras, respectively~\cite{Mazloumi:2025pmx,Capuano:2025myy,Paranjape:2026htn}. Moreover, it was also noticed~\cite{Capuano:2026pgq} they satisfy a much stronger version of cluster adjacency than has been observed~\cite{Drummond:2017ssj, Drummond:2018dfd,Golden:2019kks} in SYM theory: every letter of a given word in the symbol lies in a single cluster, a property known as total cluster compatibility.

Remarkably, Matveiakin and Rudenko~\cite{matveiakin2022cluster,rudenko2023goncharov} recently constructed and classified functions satisfying $A$-type total compatibility in terms of what they call \emph{quadrangular polylogarithm functions} ($\qli$), so named because a key step in their combinatorial construction involves summing over quadrangulations of a polygon with a certain cross-ratio attached to each quadrangle. In a companion paper~\cite{Ferro:2026oph} we presented an explicit formula (at function level, not just at symbol level), for the wavefunction coefficient associated to the $n$-site chain graph as a pure weight-$n$ linear combination of $\qli$ functions (and products thereof).

In this paper we solve the analogous problem for the $n$-gon graph. First we construct, for each $n$, a weight-$n$ function that satisfies total compatibility with respect to the $B_{2n-1}$ cluster algebra. Remarkably, the most straightforward adaptation of Matveiakin and Rudenko’s construction leads to a function $\qli_n^B$ that we prove is \emph{precisely} equal to the (loop-momentum integrand of the) wavefunction coefficient associated to the $n$-gon graph in de Sitter cosmology, with no products of lower-weight or constant terms missing. Our construction relies on the fact that a cluster algebra of type $B_m$ can be obtained by folding one of type $D_{m+1}$ by a $\mathbb{Z}_2$ symmetry, realised combinatorially by triangulations of an auxiliary $(2m+2)$-gon invariant under a $180^\circ$ rotation \cite{Fomin_2003}. Every quadrangulation of this auxiliary polygon that descends to the quotient has precisely one self-conjugate quadrangle, and we show that the associated symbol letter is the square root of the cross-ratio associated to the quadrangle, with a particular and crucial choice of sign. Consequently, we will see that the $B$-type quadrangular polylogarithms are not simply related to those of $A$-type.

The paper is organised as follows. We first review in Section \ref{sec:review} basic facts about the $B$-type cluster algebras needed in the rest of the paper. Section \ref{sec:mathematics} defines the $B$-type quadrangular polylogarithm function $\qli_n^B$ and gives explicit expressions for $n\leq 3$. Section~\ref{sec:physics} reviews the relation between the kinematics associated with the cosmological $n$-gon graph and $B$-type cluster algebras, presents the recursive differential equation satisfied by the corresponding wavefunction coefficients, and outlines the proof that it is satisfied by $\qli_n^B$, with details of the proof left to Appendix \ref{sec:Bcancel}. Section \ref{sec:examples} illustrates our result for the examples of one-, two- and three-gon graphs. In Section \ref{sec:limit2} we identify a specific kinematic limit, in which the $n$-gon wavefunction coefficient reduces to that of the $n$-site chain. We end with conclusions and open problems.

%%%%%%%%%%%%%%%%%%%%%%%%%%%%%%%%%%%%%%%%%%%%%%%%%%%%%%%%%
%%%%%%%%%%%%%%%%%%%%%%%%%%%%%%%%%%%%%%%%%%%%%%%%%%%%%%%%%

\section{Review of \texorpdfstring{$B$}{B}-type Cluster Algebras}
\label{sec:review}

In this section we review the cluster coordinates and geometric description of $B$-type cluster algebras, largely following \cite{Fomin_2003, Fomin:2001rc, fomin2021introductionclusteralgebraschapter123,fomin2021introductionclusteralgebraschapters,fomin2007clusteralgebrastriangulatedsurfaces,fomin2021introductionclusteralgebraschapter6}.
One description is given by folding a $D$-type cluster algebra by a $\mathbb{Z}_2$ automorphism (see \cite{fomin2021introductionclusteralgebraschapters} for details), which we review in Appendix \ref{app:Bcoords}, where we show that it is equivalent to the description below.

For most of the paper we shall use a second geometric description, introduced in \cite{Fomin_2003}, in which clusters of the $B_m$ algebra are in bijection with \emph{centrally symmetric triangulations} of a regular $(2m+2)$-gon $\mathbf{P}_{2m+2}$. A cluster variable corresponds to a chord and its image under a $180^\circ$ rotation: either a main diagonal, which is invariant under the rotation, or a centrally symmetric pair of non-main diagonal chords. Two cluster variables are \emph{compatible}, i.e. belong to a common cluster, if and only if the corresponding chords do not cross in the interior of the polygon. Chords that cross correspond to incompatible cluster variables, and clusters are the maximal compatible sets, i.e.~full triangulations of $\mathbf{P}_{2m+2}$. The total number of mutable cluster variables is $m(m+1)$ and the number of clusters is $\binom{2m}{m}$ \cite{fomin2021introductionclusteralgebraschapters}.

One may realise these cluster $\mathcal{A}$-coordinates, supplemented by a set of $m+1$ frozen variables, using the Pl\"ucker coordinates of a matrix representative of $\text{Gr}(2,m+1)$. Labelling the vertices of $\mathbf{P}_{2m+2}$ cyclically as $(1,\cdots,m{+}1,\bar{1},\cdots,\overline{m{+}1})$ and writing $\Delta_{ab}$ for the cluster variable attached to the chord joining $a$ and $b$, the symmetry implies that $\Delta_{\bar{a}\bar{b}}=\Delta_{ab}$ and $\Delta_{a\bar{b}}=\Delta_{\bar{a}b}$. We choose the parameterisation
\begin{equation}\label{eq:Gr}
    \begin{pmatrix}
        1 & 1 &\cdots &1\\
        z_1 & z_2 &\cdots&z_{m+1}
    \end{pmatrix}\,,
\end{equation}
in which case the appropriate correspondence is
\begin{equation}
\label{eq:deltadef}
    \Delta_{ab} = z_b - z_a\,, \quad 1\le a<b\le m+1\,.
\end{equation}
If $a$ and $b$ are adjacent then $\Delta_{ab}$ corresponds to a frozen variable; otherwise it corresponds to a mutable variable. The cluster variables $\Delta_{a\bar{b}}$ are determined as functions of these by
\begin{equation}
\begin{aligned}\label{eq:BDeltarelation}
    \Delta_{a\bar{b}}&=1-\Delta_{ab}, & 1\le a \le b \le m+1\,.
\end{aligned}
\end{equation}
All of the $\Delta_{a \bar{b}}$ variables correspond to internal chords and are mutable, except for the boundary chord $\Delta_{1\,\overline{m+1}}$ which is frozen. Setting $a=b$ in \eqref{eq:BDeltarelation} and using $\Delta_{aa}=0$, we see that $\Delta_{a\bar{a}}=1$ for $1\le a\le m+1$, namely the cluster $\mathcal{A}$-coordinates corresponding to the main diagonals are set to unity in this realisation. Note that these extra coordinates $\Delta_{a\bar b}$ \emph{cannot} be realised as Pl\"ucker coordinates of a matrix in $\text{Gr}(2,m+1)$, since they do not satisfy the Pl\"ucker relations. For future convenience we also define $\Delta_{ab}=-\Delta_{ba}$ and $\Delta_{a\bar{b}}=-\Delta_{\bar{b}a}$ for $a<b$.

%%%%%%%%%%%%%%%%%%%%%%%%%%%%%

In this presentation there are a total of $(m+1)^2$ cluster $\mathcal{A}$-coordinates, of which $m+1$ (the main diagonals) are set to 1.
$B$-type cluster algebras also admit $\mathcal{X}$-coordinates (see \cite{fomin2021introductionclusteralgebraschapter123} for details) which are certain cross-ratios of $\mathcal{A}$-coordinates.
For our discussion in Section \ref{sec:mathematics} and beyond it will be important to distinguish cross-ratios that are self-conjugate from those that are not; we will call the latter ``regular''. Given any four vertices $i,j,k,l$ of $\mathbf{P}_{2m+2}$ in cyclic order, which may be barred or unbarred but all lie inside one half of $\mathbf{P}_{2m+2}$, we define the regular cross-ratio
\begin{equation}\label{eq:normal_cross_ratios}
    \boxed{q_{ijkl}\equiv \frac{\Delta_{ij}\Delta_{kl}}{\Delta_{jk}\Delta_{li}}}
\end{equation}
which satisfies $q_{\bar i\bar j\bar k\bar l} = q_{ijkl}$ (with the understanding that $\overline{\overline{v}}=v$) and the identity
\begin{align}
    \qquad 1-q_{ijkl}=q_{ikjl}=\frac{\Delta_{ik}\Delta_{jl}}{\Delta_{kj}\Delta_{li}}\,.
\end{align}

For four vertices of the form $(a,b,\bar{a},\bar{b})$ with $a<b$, we define the self-conjugate cross-ratio
\begin{equation}\label{eq:sqrtdefinition}
\boxed{q_{ab\bar{a}\bar{b}}=\frac{\Delta_{ab}\Delta_{\bar{a}\bar{b}}}{\Delta_{b\bar{a}}\Delta_{\bar{b}a}}=w_{ab}^2\,,\qquad \text{where }\quad w_{ab}=-\frac{\Delta_{ab}}{\Delta_{a\bar{b}}}=-\frac{\Delta_{ab}}{1-\Delta_{ab}}}\,.
\end{equation}
Here, we use $\Delta_{\bar{a}\bar{b}}=\Delta_{ab}$ and $\Delta_{b\bar{a}}=\Delta_{\bar{b}a}=-\Delta_{a\bar{b}}$ to see that they are perfect squares (which is a signature of the $\mathbb{Z}_2$ folding). The choice of minus sign in the definition of $w_{ab}$ will prove crucial in the following section. 

\begin{figure}
\begin{center}
\begin{tikzpicture}[scale=1.6, every node/.style={font=\small}]
  \def\R{1.4}
  \foreach \i/\lab/\ang in {
    1/1/112.5,
    2/\bar{4}/157.5,
    3/\bar{3}/202.5,
    4/\bar{2}/247.5,
    5/\bar{1}/292.5,
    6/4/337.5,
    7/3/22.5,
    8/2/67.5
  }{
    \coordinate (v\i) at (\ang:\R);
    \node at (\ang:\R+0.25) {$\lab$};
  }
  \fill[gray!20,opacity=0.6] (v1)--(v2)--(v5)--(v6)--cycle;

  \draw[line width=1.2pt]
    (v1)--(v2)--(v3)--(v4)--(v5)--(v6)--(v7)--(v8)--cycle;

  \draw[line width=1pt] (v1)--(v5)
    node[pos=0.22,fill=white,inner sep=1pt]
    {$\Delta_{1\bar{1}}$};

  \draw[line width=1pt,red!70!black] (v2)--(v4);
  \draw[line width=1pt,red!70!black] (v6)--(v8)
    node[pos=0.5,fill=white,inner sep=1pt]
    {$\Delta_{24}$};

  \draw[line width=1pt,green!45!black] (v2)--(v5);
  \draw[line width=1pt,green!45!black] (v6)--(v1)
    node[pos=0.5,fill=white,inner sep=1pt]
    {$\Delta_{14}$};

  \foreach \i in {1,...,8}{
    \fill (v\i) circle (1.2pt);
  }
  % \fill[gray] (0,0) circle (0.8pt);
\end{tikzpicture}
\caption{A centrally symmetric triangulation of $\mathbf{P}_{8}$. The shaded quadrilateral $(1,4,\bar 1,\bar 4)$ is associated to a self-conjugate cross-ratio; the two remaining quadrilaterals $(1,2,3,4)$ and $(\bar 1,\bar 2,\bar 3,\bar4)$ are associated to a conjugate pair of regular cross-ratios.}
\label{fig:P8}
\end{center}
\end{figure}
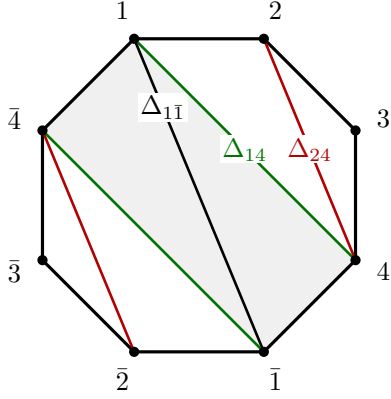

Consider the example of $B_3$. The corresponding polygon $\mathbf{P}_{8}$, with vertices labelled $(1,\ldots,4,\bar{1},\ldots,\bar{4})$, is shown in Figure~\ref{fig:P8}, which also indicates an example of a cluster, i.e.~a centrally symmetric triangulation, with internal chords $(1,\bar{1})$, $(1,4)$, $(2,4)$ and
their images, along with boundary chords $(1,2),(2,3),(3,4),(\bar{1},4)$ and their images. The associated mutable cluster variables are $\{\Delta_{1\bar{1}},\Delta_{14},\Delta_{24}\}
=\{1,\,z_4-z_1,\,z_4-z_2\}$, while the frozen variables are $\{\Delta_{12},\Delta_{23},\Delta_{34},\Delta_{\bar{1}4}\}
=\{z_2-z_1,\,z_3-z_2,\,z_4-z_3,\,1-(z_4-z_1)\}$.

We can also construct regular and self-conjugate cross-ratios for $B_3$. For example, in Figure \ref{fig:P8}, the pair of conjugate quadrilaterals (1,2,3,4) and $(\bar 1,\bar{2},\bar 3, \bar 4)$ is associated to the regular cross-ratio
\begin{equation}
    q_{1234}= q_{\bar 1\bar{2}\bar{3}\bar{4}}=-\frac{\Delta_{12}\Delta_{34}}{\Delta_{23}\Delta_{14}}\,,
\end{equation}
while the remaining quadrilateral $(1,4,\bar 1,\bar 4)$ is associated to the self-conjugate cross-ratio square root
\begin{align}
    w_{14}=-\frac{\Delta_{14}}{\Delta_{\bar14}}\,.
\end{align}

%%%%%%%%%%%%%%%%%%%%%%%%%%%%%%%%%%%%%%%%%%%%%%%%%%%%%%%%%
%%%%%%%%%%%%%%%%%%%%%%%%%%%%%%%%%%%%%%%%%%%%%%%%%%%%%%%%%

\section{\texorpdfstring{$B$}{B}-type Quadrangular Polylogarithms}\label{sec:mathematics}

In this section we define the quadrangular polylogarithm function $\text{QLi}^B_n$, which satisfies total compatibility with respect to the $B_{2n-1}$ cluster algebra.
The construction runs parallel to the $A$-type case \cite{matveiakin2022cluster,rudenko2023goncharov,Ferro:2026oph} with one major structural modification: every combinatorial object is required to be invariant under the involution $a\leftrightarrow \bar{a}$, and the resulting weighted symbols are read off from the quotient by that involution.
Note that we do not make any claim about solving the very interesting general problem of classifying and enumerating all weight-$w$ functions that satisfy total compatibility with respect to the $B_m$ cluster algebra for arbitrary $w$ and $m$.

%%%%%%%%%%%%%%%%%%%%%%%%%%%%%%%%%%%%%%%%%%%%%%%%%%%%%%%%%

\subsection{Arborification}

A centrally symmetric alternating $4n$-gon $\mathbf{P}_{4n}$ is a list
\begin{equation}
\label{eq:orderP}
    \mathbf{P}_{4n}=(p_1,p_2,\ldots, p_{2n},\overline{p}_{1},\overline{p}_{2},\ldots,\overline{p}_{2n})
\end{equation}
of $4n$ integers, with $p_1$ odd, labelling the vertices of the $4n$-gon in cyclic order, alternating between even and odd, and invariant under the involution $\sigma$ that sends each vertex $p_i$ to the vertex $\overline{p}_i$ $2n$ positions away. Throughout, we use the convention $\overline{\overline{v}}=v$. We write $\pi(v)\in\{1,\ldots,4n\}$ for the position of a vertex, so that $\pi(\bar{v})=\pi(v)+2n\ (\text{mod } 4n)$. Because $2n$ is even, $\sigma$ preserves the parity of positions. An alternating subpolygon of $\mathbf{P}_{4n}$ (including $\mathbf{P}_{4n}$ itself) is called even (odd) according to the parity of the smallest position among its vertices. Therefore the involution $\sigma$ also preserves the parity of an alternating polygon. Along with the preservation of parity of each position, this ensures that our construction descends to the quotient.

We define a \emph{partial quadrangulation} of $\mathbf{P}_{4n}$ to consist of a self-conjugate alternating 4$k$-gon,
\begin{equation}\label{eq:oddcent}
    \mathbf{S}_{4k}=(s_1,\ldots,s_{2k},\bar{s}_1,\ldots,\bar{s}_{2k})\,,\qquad 1\leq k\leq n\,,
\end{equation}
which we call the \emph{central polygon}, together with full quadrangulations of the connected components of $\mathbf{P}_{4n}\setminus \mathbf{S}_{4k}$ invariant under $\sigma$. Note that we always choose the cyclic representative of $\mathbf{S}_{4k}$ for which $\mathbf{S}_{4k}(1)$ is odd. We use $\mathcal{Q}^\sigma(\mathbf{P}_{4n})$ to denote the set of \emph{partial quadrangulations} of $\mathbf{P}_{4n}$. 

The dual graph $\widetilde{Q}$, in which the central polygon corresponds to a node, is therefore a tree carrying a $\mathbb{Z}_2$ symmetry with exactly one invariant node, the central node, and the quotient
\begin{equation}
    \hat{Q}=\widetilde{Q}/\sigma
\end{equation}
is a tree on $(n-k+1)$ nodes rooted at the central node. This is the point at which the $B$-type construction departs from the $A$-type one of~\cite{matveiakin2022cluster,rudenko2023goncharov,Ferro:2026oph}. For chain graphs, the root of $\widetilde{Q}$ has to be selected by hand by choosing the node containing a distinguished boundary edge; here the involution supplies a root canonically, and no boundary marking is needed. The root of $\hat{Q}$ has degree\footnote{The degree of a node is the number of edges incident on it.} at most $2k$, since the $4k$ sides of the central polygon are identified in pairs by $\sigma$, while every other node has degree at most four.

Next we associate to every $\mathbf{P}_{4n}$ a weighted symbol $T_\mathbf{P}^B$ in the Connes-Kreimer Hopf algebra of rooted trees. Each symbol letter includes a variable $\varphi$ and a positive integer weight $m$, and is written as $[\varphi,m]$. $T_{\mathbf{P}}^B$ is then constructed by the following rules:
\begin{enumerate}
    \item Draw a partial quadrangulation $Q\in\mathcal{Q}^\sigma(\mathbf{P}_{4n})$ together with the quotient tree $\hat{Q}$. 
    \item 
    Colour each non-root node of $\hat{Q}$ white (black) if the quadrilateral it represents is even (odd). This is well-defined because $\sigma$ preserves the parity of polygons.
    \item Mark the root node, the image of the central polygon $\mathbf{S}_{4k}$, with a square.
    \item Draw an arrow on each edge of $\hat{Q}$ pointing away from the root.
    \item Each non-root node $R=(a,b,c,d)$ is labelled such that $a,b,c,d$ all lie in the same half of the polygon and are ordered in the sense of \eqref{eq:orderP}. $R$ is further labelled by the signed weighted letter
    \begin{equation}\label{eq:mainepsxQ}
        \begin{cases}
            +[q_{abcd},1]\,,&\pi(a) \text{ even},\\
            -[1/q_{abcd},1]\,,&\pi(a) \text{ odd},
        \end{cases}
    \end{equation}
    where $q_{abcd}$ is the cross-ratio defined in \eqref{eq:normal_cross_ratios}. This is unambiguous on the quotient because conjugate quadrilaterals carry equal cross-ratios, $q_{\bar{a}\bar{b}\bar{c}\bar{d}}=q_{abcd}$.

    \item For the root node, which is the image of the central $4k$-gon $\mathbf{S}_{4k}$, note that before evaluating the generalised cross-ratio $\text{cr}^B(\mathbf{S}_{4k})$, we cyclically rotate $\mathbf{S}_{4k}$, if necessary, so that $\mathbf{S}_{4k}(1)$ is odd. Then we label it by the weight-$k$ letter $[\text{cr}^B(\mathbf{S}_{4k}),k]$, with
    \begin{equation}
    \label{eq:crdefBroot}
        \text{cr}^B(\mathbf{S}_{4k})=(-1)^{k+1}\frac{\prod_{j=1}^{k}\Delta_{\mathbf{S}_{4k}(2j)\mathbf{S}_{4k}(2j+1)}}{\prod_{j=1}^{k}\Delta_{\mathbf{S}_{4k}(2j+1)\mathbf{S}_{4k}(2j+2)}}\,.
    \end{equation}
    Here the indices are understood in the cyclic ordering of $\mathbf{S}_{4k}$.
    
    \item Denote the connected components in one half of the complement $\mathbf{P}_{4n}\backslash \mathbf{S}_{4k}$ by $\{S_i\}_{i\geq1}$. Define $t_{S_i}^A$ to be the weighted symbol of the rooted dual tree associated with the chosen quadrangulation of each $S_i$ recursively as follows. Given a root $R=(a,b,c,d)$ and the weighted symbols of its three descendant subtrees $t_1,t_2,t_3$, we apply the following two rules for the different parities of $a$\footnote{For conventional non-weighted symbols, $[1/\varphi]=-[\varphi]$, which would make the odd case look like a rewriting of the even one. However, it is important to note that as weighted symbols they are distinct formal expressions, and they will be promoted to particular MPL functions in Section~\ref{sec:threepointthree}.}:
    \begin{align}\label{eq:Aarbe}
    \raisebox{-5pt}{\begin{tikzpicture}[baseline=(current bounding box.center), scale=1.4]
    \draw[arb] (0,0) -- (-0.8,0);
    \draw[arb] (0,0) -- (0,0.8);
    \draw[arb] (0,0) -- (0.8,0);
    \filldraw[fill=white] (0,0) circle (2pt) node[below=6pt] {$[q_{abcd},1]$};
    \filldraw[fill=gray] (-0.8,0) circle (2pt) node[left=4pt] {$t_1$};
    \filldraw[fill=gray] (0,0.8) circle (2pt) node[above=4pt] {$t_2$};
    \filldraw[fill=gray] (0.8,0) circle (2pt) node[right=4pt] {$t_3$};
    \end{tikzpicture}}\hspace{-2mm}
    &\hspace{-2mm}= (t_1\star t_2\star t_3)\otimes[q_{abcd},1]+(t_1\star t_2\star t_3)\cdot[q_{abcd},1]\,,~ \pi(a) \text{ even},
    \\[4pt]\label{eq:Aarbo}
    \raisebox{-5pt}{\begin{tikzpicture}[baseline=(current bounding box.center), scale=1.4]
    \draw[arb] (0,0) -- (-0.8,0);
    \draw[arb] (0,0) -- (0,0.8);
    \draw[arb] (0,0) -- (0.8,0);
    \filldraw[fill=black] (0,0) circle (2pt) node[below=6pt] {$[1/q_{abcd},1]$};
    \filldraw[fill=gray] (-0.8,0) circle (2pt) node[left=4pt] {$t_1$};
    \filldraw[fill=gray] (0,0.8) circle (2pt) node[above=4pt] {$t_2$};
    \filldraw[fill=gray] (0.8,0) circle (2pt) node[right=4pt] {$t_3$};
    \end{tikzpicture}}
    &\hspace{-2mm}= -(t_1\star t_2\star t_3)\otimes[1/q_{abcd},1]\,,~ \pi(a) \text{ odd}.
    \end{align}
     Empty subtrees are replaced by the empty word $1$. The contribution of a particular partial quadrangulation of a $4n$-gon is
    \begin{equation}\label{eq:tQB}
        t^B_Q=(-1)^n\left(\underset{i\geq1}{\bigstar}t_{S_i}^A\right)\otimes[\text{cr}^B(\mathbf{S}_{4k}),k]\,,
    \end{equation}
    where $t_{S_i}^A$ is as defined by \eqref{eq:Aarbe}--\eqref{eq:Aarbo}.
    \item Finally, sum over all partial quadrangulations to get the weighted symbol associated to the alternating polygon $\mathbf{P}_{4n}$:
    \begin{equation}\label{eq:Tarpp}
        T_{\mathbf{P}}^B=\sum_{Q\in\mathcal{Q}^\sigma(\mathbf{P}_{4n})}t_Q^B\,.
    \end{equation}
\end{enumerate}

The products appearing in step~7 are the same as in~\cite{Ferro:2026oph}: the \emph{dot product}
\begin{equation}
\label{eq:BproductdefB}
    \bigl( [\varphi_1,m_1]\otimes\cdots\otimes[\varphi_r,m_r] \bigr) \cdot [\varphi,m] 
    = [\varphi_1,m_1]\otimes\cdots\otimes [\varphi_r\varphi,m_r+m]\,,
\end{equation}
and the \emph{star product}, or \emph{quasi-shuffle product}, defined recursively by
\begin{multline}
\label{eq:BstardefB}([\varphi_1,m_1]\otimes\omega_1)\star([\varphi_2,m_2]\otimes\omega_2)
    =[\varphi_1,m_1]\otimes\bigl((\omega_1)\star([\varphi_2,m_2]\otimes\omega_2)\bigr)\\
    +[\varphi_2,m_2]\otimes\bigl(([\varphi_1,m_1]\otimes\omega_1)\star(\omega_2)\bigr)
    +\bigl([\varphi_1,m_1]\cdot[\varphi_2,m_2]\bigr)\otimes(\omega_1\star\omega_2)\,,
\end{multline}
with $1\star\omega=\omega\star 1=\omega$ and $1 \cdot [\varphi,m]=0$, extended linearly to sums of words.

The arborification of each complementary polygon uses exactly the same weighted letters, rooted-tree combinatorics, dot product, and quasi-shuffle product as the arborification used for the $A$-type functions, except for two changes. First, the ordering of the root letter and the quasi-shuffle product of its descendant words is reversed. Second, the parity of the contraction term is reversed: the additional contraction (contributed by the dot product) is present at an even root instead of an odd root.

The $B$-type construction adds one further ingredient. The self-conjugate central polygon $\mathbf{S}_{4k}$ is not necessarily a quadrilateral, and enters the arborification as a weight-$k$ letter $[\text{cr}^B(\mathbf{S}_{4k}),k]$. In our physical application, the advantage of this new arborification rule is that every wavefunction coefficient of a polygon graph can be represented by a single quadrangular polylogarithm under the modified definition, which is considerably simpler than the corresponding representation for chain graphs.

%%%%%%%%%%%%%%%%%%%%%%%%%%%%%%%%%%%%%%%%%%%%%%%%%%%%%%%%%

\subsection{Examples of Partial Quadrangulations and Their Associated Words}\label{sec:tarpBexample}
We now look at a few examples, namely $n=1,2,3$. In the figures in this section, we draw the full dual tree $\widetilde{Q}$ rather than the quotient tree $\hat{Q}$ used in the arborification in order to make the $\mathbb{Z}_2$ symmetry manifest (and so each conjugate pair of nodes in $\widetilde{Q}$ represents a single node of $\hat{Q}$). 

The first case is $n=1$, where $\mathbf{P}_{4}=(1,2,\bar 1,\bar 2)$ is a 4-gon. Here, we have a single partial quadrangulation, consisting of the central polygon alone. Its quotient tree is a single root node with no subtrees,
\begin{equation}
\label{eq:BsquareB}
    T^B_{(1,2,\bar 1,\bar 2)}=
    \begin{tikzpicture}[baseline=(current bounding box.center), scale=0.85]
      \foreach \i/\lab in {0/1, 1/2, 2/\bar 1, 3/\bar 2} {
        \coordinate (V\i) at ({cos(90-\i*90+45)},{sin(90-\i*90+45)});
        \node at ({1.3*cos(90-\i*90+45)},{1.3*sin(90-\i*90+45)}) {$\lab$};
      }
      \draw (V0)--(V1)--(V2)--(V3)--cycle;
      \draw (-3pt,-3pt) rectangle (3pt,3pt);
    \end{tikzpicture}
    =-\bigl[1/w_{12},1\bigr]\,, 
\end{equation}
with $w_{ab}$ defined in \eqref{eq:sqrtdefinition}. 

\begin{figure}
\centering
\begin{tabular}{ccc}
% ---------------- Q1 ----------------
\begin{tikzpicture}[scale=1.35,baseline=(current bounding box.center)]
  \octbase
  \fill[gray!18,opacity=0.6] (v1)--(v2)--(v5)--(v6)--cycle;
  \draw[line width=0.9pt] (v2)--(v5);
  \draw[line width=0.9pt] (v6)--(v1);
  \draw[arb] (0,0)--(0.653,0.000);
  \draw[arb] (0,0)--(-0.653,-0.000);
  \filldraw[fill=white] (0.653,0.000) circle (2pt);
  \filldraw[fill=white] (-0.653,-0.000) circle (2pt);
  \draw (-3.2pt,-3.2pt) rectangle (3.2pt,3.2pt);
\end{tikzpicture}
&
\begin{tikzpicture}[scale=1.35,baseline=(current bounding box.center)]
  \octbase
  \fill[gray!18,opacity=0.6] (v2)--(v3)--(v6)--(v7)--cycle;
  \draw[line width=0.9pt] (v3)--(v6);
  \draw[line width=0.9pt] (v7)--(v2);
  \draw[arb] (0,0)--(0.462,-0.462);
  \draw[arb] (0,0)--(-0.462,0.462);
  \fill (0.462,-0.462) circle (2pt);
  \fill (-0.462,0.462) circle (2pt);
  \draw (-3.2pt,-3.2pt) rectangle (3.2pt,3.2pt);
\end{tikzpicture}
&
\begin{tikzpicture}[scale=1.35,baseline=(current bounding box.center)]
  \octbase
  \fill[gray!18,opacity=0.6] (v3)--(v4)--(v7)--(v8)--cycle;
  \draw[line width=0.9pt] (v4)--(v7);
  \draw[line width=0.9pt] (v8)--(v3);
  \draw[arb] (0,0)--(0.000,-0.653);
  \draw[arb] (0,0)--(0.000,0.653);
  \filldraw[fill=white] (0.000,-0.653) circle (2pt);
  \filldraw[fill=white] (0.000,0.653) circle (2pt);
  \draw (-3.2pt,-3.2pt) rectangle (3.2pt,3.2pt);
\end{tikzpicture}
\\[4pt]
$Q^{(1)}$ & $Q^{(2)}$ & $Q^{(3)}$ %& $Q^{(4)}$ & $Q^{(5)}$
\end{tabular}
\begin{tabular}{cc}
    \begin{tikzpicture}[scale=1.35,baseline=(current bounding box.center)]
    \octbase
    \fill[gray!18,opacity=0.6] (v1)--(v4)--(v5)--(v8)--cycle;
    \draw[line width=0.9pt] (v1)--(v4);
    \draw[line width=0.9pt] (v5)--(v8);
    \draw[arb] (0,0)--(0.462,0.462);
    \draw[arb] (0,0)--(-0.462,-0.462);
    \fill (0.462,0.462) circle (2pt);
    \fill (-0.462,-0.462) circle (2pt);
    \draw (-3.2pt,-3.2pt) rectangle (3.2pt,3.2pt);
    \end{tikzpicture}
    &
    \begin{tikzpicture}[scale=1.35,baseline=(current bounding box.center)]
    \octbase
    \fill[gray!18,opacity=0.6] (v1)--(v2)--(v3)--(v4)--(v5)--(v6)--(v7)--(v8)--cycle;
    \draw (-3.2pt,-3.2pt) rectangle (3.2pt,3.2pt);
    \end{tikzpicture}
    \\[4pt]
    $Q^{(4)}$ & $Q^{(5)}$
\end{tabular}
\caption{The five partial quadrangulations of the octagon, $n=2$. For the first four,
the shaded region is the central, self-conjugate quadrilateral, whose dual node is marked
with a square; the two unshaded quadrilaterals form a conjugate pair and are identified
in the quotient. In the last picture the entire octagon is the central polygon, so
the quotient tree consists only of its root. Ordinary nodes are white for even
quadrilaterals and black for odd ones. The five pictures correspond term by term to the
five terms of~(\ref{eq:BoctagonB}).}
\label{fig:Boctagon}
\end{figure}
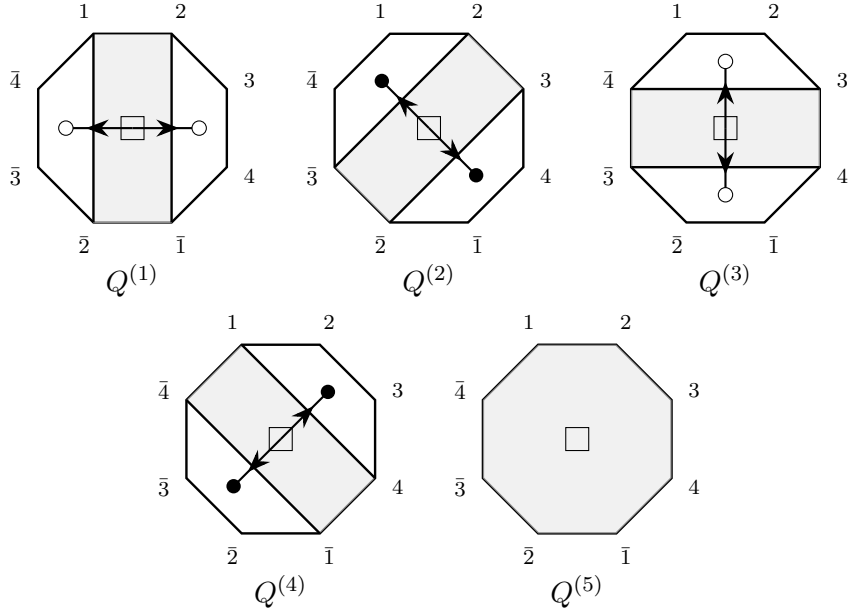

The first non-trivial case is $n=2$, the octagon $\mathbf{P}_{8}=(1,2,3,4,\bar 1,\bar 2,\bar 3,\bar 4)$.
There are four partial quadrangulations whose central polygon is a quadrilateral, together
with one for which the whole octagon is the central polygon (see Figure~\ref{fig:Boctagon}):
\begin{equation}
\label{eq:BoctagonlistB}
\begin{aligned}
    Q^{(1)}&:\ (1,2,\bar 1,\bar 2)\ \text{central},\quad (2,3,4,\bar 1)\,, &\quad
    Q^{(2)}&:\ (2,3,\bar 2,\bar 3)\ \text{central},\quad (3,4,\bar 1,\bar 2)\,,\\
    Q^{(3)}&:\ (3,4,\bar 3,\bar 4)\ \text{central},\quad (4,\bar 1,\bar 2,\bar 3)\,, &\quad
    Q^{(4)}&:\ (1,4,\bar 1,\bar 4)\ \text{central},\quad (1,2,3,4)\,,\\
    Q^{(5)}&:\ (1,2,3,4,\bar 1,\bar 2,\bar 3,\bar 4)\ \text{central}. &&
\end{aligned}
\end{equation}

The first four quotient trees consist of a root with a single child, while the fifth
consists of the root alone. The children of $Q^{(1)},Q^{(3)}$ are even and those of
$Q^{(2)},Q^{(4)}$ are odd. Applying the rules above gives
\begin{equation}
\label{eq:BoctagonB}
\boxed{
\begin{aligned}
    T^B_{(1,2,3,4,\bar1,\bar2,\bar3,\bar4)}
    = &+ [q_{234\bar1},1] \otimes [1/w_{12},1] - [1/q_{34\bar1\bar2},1] \otimes [w_{23},1]\\
    &+ [q_{4\bar1\bar2\bar3},1] \otimes [1/w_{34},1] - [1/q_{1234},1] \otimes [1/w_{14},1] \\
    & + [w_{23}/q_{34\bar1\bar2},2] \,. 
\end{aligned}}
\end{equation}
The five terms of~(\ref{eq:BoctagonB}) are in bijection with the five partial
quadrangulations of the octagon. In all five terms the total weight is two.

\begin{figure}
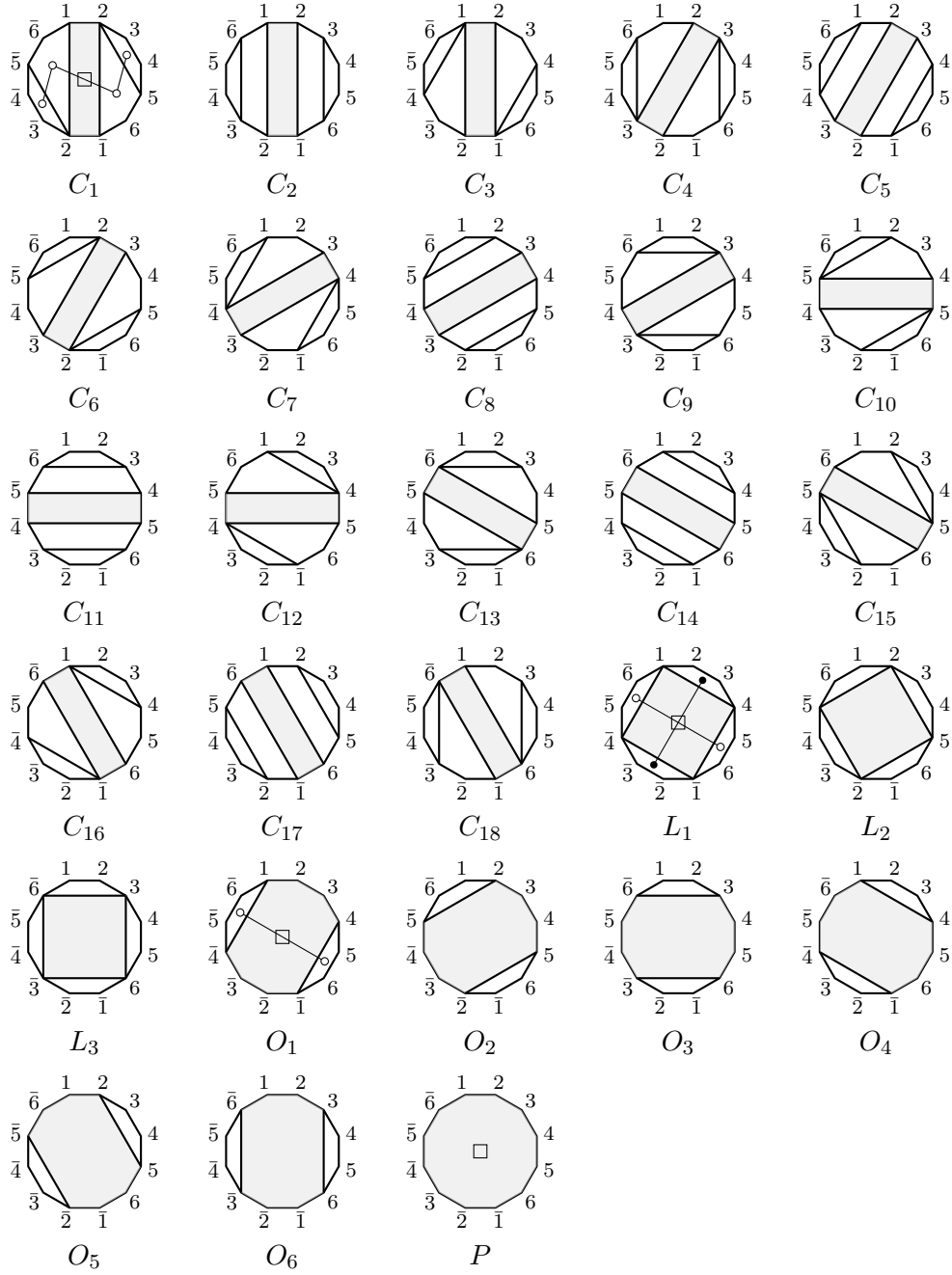

\centering
\include{Figures/Fig4}
\caption{The 28 partial quadrangulations of the dodecagon, $n=3$. The shaded region is the central polygon and carries the root node; the unshaded quadrilaterals form conjugate pairs, identified in the quotient tree. The first 18 $C_i$ have a central quadrilateral of thin type, e.g. $(1,2,\bar 1,\bar 2)$. The next 3 $L_i$ have a central quadrilateral of thick type, e.g. $(1,4,\bar 1,\bar 4)$. Moving on to central octagons, e.g.~$(1,2,3,4,\bar1, \bar2, \bar3, \bar4)$, we have the $O_i$. Finally, $P$ is the whole polygon. Ordinary nodes are white for even quadrilaterals and black for odd ones. We only show the dual trees of the partial quadrangulations discussed.}
\label{fig:Bdodecagon}
\end{figure}

For $n=3$ the polygon is the dodecagon $\mathbf{P}_{12}=(1,\ldots,6,\bar 1,\ldots,\bar 6)$. There are three possible types of central polygon: a quadrilateral, an octagon, or the entire dodecagon itself. A central quadrilateral is then one of two types,
\begin{equation}
\label{eq:BcentralCL}
\begin{aligned}
    \text{thin:}&\quad (p,p+1,\overline{p},\overline{p+1})\,, &\quad& p=1,\ldots,6\,,\\
    \text{thick:}&\quad (p,p+3,\overline{p},\overline{p+3})\,, &\quad& p=1,2,3\,,
\end{aligned}
\end{equation}
where $7$ and $\bar 7$ are identified with $\bar 1$ and $1$. For the thin central quadrilateral, two of the four sides are the boundary edges and the other two sides are shared with a hexagon. Each orientation of the thin central quadrilateral contributes three partial quadrangulations corresponding to one of the three quadrangulations of the hexagon. These are the 18 partial quadrangulations $C_{i}$  illustrated in Figure \ref{fig:Bdodecagon}. For the thick central quadrilateral, all four sides are shared with a quadrilateral, and each orientation contributes one partial quadrangulation. These are the 3 partial quadrangulations $L_i$ given in Figure \ref{fig:Bdodecagon} and the first example of a forked quotient tree. Finally, there are six partial quadrangulations whose central polygon is an octagon, and one whose central polygon is a dodecagon. These partial quadrangulations are denoted as $O_i$ and $P$, respectively, in Figure \ref{fig:Bdodecagon}.

The full weighted symbol for the dodecagon is the sum over all 28 partial quadrangulations,
\begin{equation}
\label{eq:BdodecagonB}
\boxed{
    T^B_{(1,\ldots,6,\bar 1,\ldots,\bar 6)}
    =\sum_{i=1}^{18}t^B_{C_i}\;+\;\sum_{i=1}^{3}t^B_{L_i}
    +\sum_{i=1}^{6}t^B_{O_i}\;+\;t^B_P\,.
}
\end{equation}
We provide one example of $t^B_Q$ for each type of central polygon below:
\begin{equation}
\begin{aligned}
    t^B_{C_1} &=-[q_{2345},1]\otimes[q_{256\bar1},1]\otimes\left[1/w_{12},1\right]-[q_{2345}q_{256\bar1},2]\otimes\left[1/w_{12},1\right],\\
    t^B_{L_1} &=[q_{456\bar1},1]\otimes\left[1/q_{1234},1\right]\otimes\left[1/w_{14},1\right]\\
    &\qquad\qquad+\left[1/q_{1234},1\right]\otimes[q_{456\bar1},1]\otimes\left[1/w_{14},1\right]+\left[q_{456\bar1}/q_{1234},2\right]\otimes\left[1/w_{14},1\right],\\
    t^B_{O_1}&=-[q_{456\bar1},1]\otimes\left[w_{23}/q_{34\bar1\bar2},2\right],\\
    t^B_P &=-[w_{23}/(q_{3456}q_{36\bar1\bar2}),3]\,.
\end{aligned}
\end{equation}
Each term in~(\ref{eq:BdodecagonB}) has weight three.

%%%%%%%%%%%%%%%%%%%%%%%%%%%%%%%%%%%%%%%%%%%%%%%%%%%%%%%%%

\subsection{\texorpdfstring{$B$}{B}-type Quadrangular Polylogarithms}
\label{sec:threepointthree}

The passage from weighted symbols to functions is simpler here than in the $A$-type case. In the $A$-type case, a $2n$-gon is decomposed into $n-1$ quadrilaterals, so the associated symbol has weight $n-1$ and one must apply the weight-raising operator $\text{Li}_k^R$ with $k\geq 1$ to reach the functions of relevance for our physical application. Here, a partial quadrangulation with a central $4k$-gon carries one central letter of weight $k$, while the fully quadrangulated complementary polygons contribute a symbol word of weight $n-k$. Therefore every term in $T_{\mathbf{P}_{4n}}^B$ has total weight $n$. No weight raising is needed, and it is sufficient to define
\begin{equation}\label{eq:Li0R}
    \text{Li}([\varphi_1,m_1]\otimes\ldots\otimes[\varphi_d,m_d]) = \text{Li}_{m_1,\ldots,m_d}(\varphi_1,\ldots,\varphi_d)\,,
\end{equation}
extended to sums of words by $\text{Li}(W_1\pm W_2\pm\ldots) = \text{Li}(W_1) \pm \text{Li}(W_2) \pm \ldots$, in terms of the multiple polylogarithms,
\begin{equation}\label{eq:Li-convention}
 \Li_{m_1,\ldots,m_d}(z_1,\ldots,z_d)
 =\sum_{0<n_1<\cdots<n_d}
 \frac{z_1^{n_1}\cdots z_d^{n_d}}
 {n_1^{m_1}\cdots n_d^{m_d}}\,,
\end{equation}
with the argument ordering convention of \cite{Ferro:2026oph}. 

We define the weight-$n$ $B$-type quadrangular polylogarithm associated with a centrally symmetric alternating $4n$-gon $\mathbf{P}_{4n}$ by
\begin{equation}\label{eq:Bpolylogs}
    \mathrm{QLi}^{B}_n(a_1,\ldots,a_{2n},\bar{a}_1,\ldots,\bar{a}_{2n})
    =\text{Li}\left(T^B_{(1,\ldots,2n,\overline{1},\ldots,\overline{2n})}\right)\bigg|_{i \to a_i}\,.
\end{equation}
Unlike the $A$-type case there is no free index: the weight is fixed by the size of the polygon. This is the $B$-type counterpart of the statement that the $A$-type function $F_n$ in \cite{Ferro:2026oph}, attached to a \emph{rooted} $2n$-gon, has weight $n$ with $n-1$ quadrilaterals in each quadrangulation. The marked root edge, which breaks cyclic invariance in the chain case, is replaced here by the central polygon, which is canonically determined for a given quadrangulation by the involution; the $B$-type building blocks are consequently invariant under rotation by two vertices without any additional marking.

Note that we interpret \eqref{eq:Bpolylogs} as defining $\QLi_n^{B}$ as an exact function, using~(\ref{eq:Li-convention}), rather than an equivalence class of functions modulo products of functions of lower weight. Throughout we write the arguments in the form $(a_1,\ldots,a_{2n},\bar{a}_{1},\ldots,\bar{a}_{2n})$, so that the pairing under the involution is manifest.

Now we claim that the functions QLi$^B_n$ live up to their name: they are $B$-type quadrangular polylogarithms, which means that their symbol letters are cluster variables and their symbols satisfy total compatibility with respect to the $B_{2n-1}$ cluster algebra. This claim will be proven in the next section by demonstrating that these functions are identically equal to $n$-gon wavefunction coefficients in de Sitter space, which are known to satisfy these properties by the arguments of \cite{Paranjape:2026htn} and~\cite{Capuano:2026pgq}. Total compatibility may also be seen more directly by inspecting the differential equation \eqref{eq:QLiB_differential_eq} that $\qli^B_n$ satisfies: taking the $(n-1,1)$ coproduct component splits the $4n$-gon into two sub-polygons that are mutually compatible, recursively building a function that manifestly satisfies total compatibility.

\subsection{Examples of \texorpdfstring{$\QLi^B_n$}{QLiB} Functions}
\label{subsubsec:Qliexamples}
For $n=1$, the weighted symbol is the single letter \eqref{eq:BsquareB}, and \eqref{eq:Li0R} gives
\begin{equation}
    \label{eq:QLiB1} \mathrm{QLi}^{B}_1(a,b,\bar{a},\bar{b})=-\text{Li}_1(1/w_{ab})\,.
\end{equation}
This is the $B$-type counterpart of the weight-one $A$-type quadrangular polylogarithm, but with the square root $w_{ab}$ of the self-conjugate cross-ratio in place of an ordinary cross-ratio.

For $n=2$, we apply the map \eqref{eq:Li0R} to the octagon weighted symbol \eqref{eq:BoctagonB}, obtaining
\begin{align}
\label{eq:BQLi2}
    \mathrm{QLi}^{B}_2(a,b,c,d,\bar a,\bar b,\bar c,\bar d)
    = & +\text{Li}_{1,1}\bigl(q_{bcd\bar a},\,1/w_{ab}\bigr) -\text{Li}_{1,1}\bigl(1/q_{cd\bar a\bar b},\,w_{bc}\bigr) \nonumber \\
    & + \text{Li}_{1,1}\bigl(q_{d\bar a\bar b\bar c},\,1/w_{cd}\bigr) - \text{Li}_{1,1}\bigl(1/q_{abcd},\,1/w_{ad}\bigr) \nonumber \\
    & +\text{Li}_{2}\bigl(w_{bc}/q_{cd\bar a\bar b}\bigr) \,.
\end{align}
The five terms are the images under $\text{Li}$ of the five partial quadrangulations of the octagon displayed in Figure~\ref{fig:Boctagon}. 

For $n=3$, $\qli^B_3$ is given by the sum over the 28 partial quadrangulations shown in Figure \ref{fig:Bdodecagon},
\begin{equation}
    \label{eq:QliB3}
    \qli^B_3(a,\dots, f,\bar{a},\dots,\bar{f}) = \sum_{i=1}^{18}\li(t^B_{C_i})\;+\;\sum_{i=1}^{3}\li(t^B_{L_i})
    +\sum_{i=1}^{6}\li(t^B_{O_i})\;+\;\li(t^B_P)\,.
\end{equation}
Here we provide one example of each type of term in \eqref{eq:QliB3}:
\begin{equation}
\begin{aligned}
    \li(t_{C_1}^B) &= -\li_{1,1,1}(q_{bcde},q_{bef\bar a},1/w_{ab})-\li_{2,1}(q_{bcde}q_{bef\bar a},1/w_{ab})\,,\\
    \li(t^B_{L_1}) &= +\li_{1,1,1}(q_{def\bar a},1/q_{abcd},1/w_{ad})\\
    &\qquad\qquad+\li_{1,1,1}(1/q_{abcd},q_{def\bar a},1/w_{ad})+\li_{2,1}(q_{def\bar a}/q_{abcd},1/w_{ad})\,,\\
    \li(t^B_{O_1})&=-\li_{1,2}(q_{def\bar a},w_{bc}/q_{cd\bar a\bar b})\,,\\
    \li(t^B_P) &= -\li_3(w_{bc}/(q_{cdef}q_{cf\bar a\bar b}))\,.
\end{aligned}
\end{equation}
Note that had we chosen the opposite sign for the square root in \eqref{eq:sqrtdefinition}, we would have ended up with multiple polylogarithms with symbol letters that are not $B$-type cluster variables. Specifically, the other sign choice would have led to functions of the form $\qli^B_1(a,b,\bar a,\bar b)=-\li_1(\Delta_{a\bar{b}}/\Delta_{ab})=\log((2\Delta_{ab}-1)/\Delta_{ab})$ but $(2\Delta_{ab}-1)$ is not a cluster variable.
%%%%%%%%%%%%%%%%%%%%%%%%%%%%%%%%%%%%%%%%%%%%%%%%%%%%%%%%%
%%%%%%%%%%%%%%%%%%%%%%%%%%%%%%%%%%%%%%%%%%%%%%%%%%%%%%%%%

\section{Polygon Graphs in dS Cosmology}
\label{sec:physics}

In this section we present our main result: the wavefunction coefficient\footnote{We always mean the \emph{integrand} of the wavefunction coefficient, since we do not carry out the integral over the loop momenta.} for the $n$-gon graph, which we will denote by $\psi_n^\circ$, is the $B$-type quadrangular polylogarithm of weight $n$ associated to a $4n$-gon. The statement is the polygon counterpart of the chain result of \cite{Ferro:2026oph}, but is structurally much simpler. Whereas the wavefunction coefficient of the $n$-site chain graph, denoted $\psi_n$, was expressed there as a linear combination of three permutations of a function called $F$, each given by a full-weight $\QLi$ function plus a sum of products of lower-weight pieces, the wavefunction coefficient $\psi_n^\circ$ is precisely the single weight-$n$ function $\QLi^B_n$ defined in the previous section. We begin with the kinematic map between cosmological kinematics and cluster variables in Section \ref{subsec:kinematics}. We present the differential equation satisfied by these wavefunction coefficients in Section \ref{sec:recursion}, followed by its solution in Section \ref{sec:solution}.

%%%%%%%%%%%%%%%%%%%%%%%%%%%%%%%%%%%%%%%%%%%%%%%%%%%%%%%%%

\subsection{Cosmological Variables}
\label{subsec:kinematics}
The kinematics of the $n$-gon graph are specified by $X_i$, the norm of total momentum flowing into site $i$, and $Y_{i\,i+1}$ the norm of the momentum along the edge between consecutive sites $i$ and $i+1$:
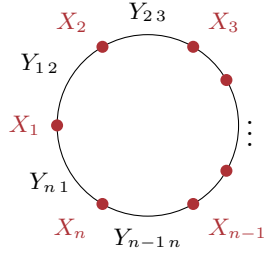
\begin{center}
  \begin{tikzpicture}[vertex/.style={circle,fill=Maroon,inner sep=1.6pt}, scale=1.5]
  \def\r{0.8}
  \draw (0,0) circle (\r);  \node[vertex,label=left:\textcolor{Maroon}{\footnotesize$X_1$}]        at (180:\r) {};
  \node[vertex,label=above left:\textcolor{Maroon}{\footnotesize$X_2$}] at (120:\r) {};
  \node[vertex,label=above right:\textcolor{Maroon}{\footnotesize$X_3$}] at (60:\r) {};
  \node[vertex,label=below left:\textcolor{Maroon}{\footnotesize$X_n$}] at (-120:\r) {};
  \node[vertex,label=below right:\textcolor{Maroon}{\footnotesize$X_{n-1}$}] at (-60:\r) {};
  \node[vertex] at (30:\r) {};
  \node[vertex] at (-30:\r) {};
  \node at (0:{\r+0.1}) {$\vdots$};
  \node at (150:{\r+0.3}) {\footnotesize$Y_{1\,2}$};
  \node at (90:{\r+0.2}) {\footnotesize$Y_{2\,3}$};
  \node at (-150:{\r+0.2}) {\footnotesize$Y_{n\,1}$};
  \node at (-90:{\r+0.2}) {\footnotesize$Y_{n{-}1\,n}$};
\end{tikzpicture}
\end{center}
The symbol alphabet of $\psi_n^\circ$ can be determined via differential equations \cite{He:2024olr,Baumann:2024mvm} and is given by the collection of letters
\begin{align}
\label{eq:xpmxpmdef}
X_{ij}^{\pm \pm} = \sum_{a=i}^{j}X_a \pm Y_{i-1\,i} \pm Y_{j\,j+1}\,,
\end{align}
where $1\leq i ,j\leq n$ and the sum over sites $a$ runs clockwise along the loop from $i$ to $j$. The mapping between the symbol alphabet of $\psi_n^\circ$ and $B_{2n-1}$ cluster variables was found in \cite{Paranjape:2026htn}. Additionally, \cite{Paranjape:2026htn} showed that its symbol has the property of cluster adjacency: adjacent symbol letters are compatible cluster variables. Remarkably, \emph{total} cluster compatibility (i.e.~all letters in a word belong to a single cluster) that was shown to exist for chain graphs \cite{Capuano:2026pgq} extends (via similar arguments) to the polygon graph contributions that we consider in this paper.

The map from kinematics to $\Delta_{ab}$ is fixed by specifying consecutive differences,
\begin{equation}
\begin{aligned}\label{eq:Delta_from_XY}
&\Delta_{2a-1\,2a}=\Delta_{\overline{2a-1}\,\overline{2a}}=\frac{X_a+Y_{a-1\,a}+Y_{a\,a+1}}{X}\,,\quad 1\leq a\leq n\,,\\
&\Delta_{2a\,2a+1}=\Delta_{\overline{2a}\,\overline{2a+1}}=\frac{-2Y_{a\,a+1}}{X}\,,\quad 1\leq a<n\,,\\
&\Delta_{1\,\overline{2n}}=\Delta_{\overline{1}\,2n}=\frac{-2Y_{n\,1}}{X}\,,
\end{aligned}
\end{equation}
where $X=\sum_{i=1}^n X_i$ and the site labels are understood mod $n$, so that $Y_{n\,n+1}=Y_{n\,1}$. 
The remaining variables follow from 
\begin{equation}
\begin{aligned}
    \label{eq:Delta_fromXY2}&\Delta_{ab}=\Delta_{\overline{a}\overline{b}}=\Delta_{a\,a+1}+\ldots+\Delta_{b-1\,b},\quad 1\leq a\leq b\leq 2n\,,\\
    &\Delta_{a\overline{b}}=\Delta_{\overline{a}b}=\Delta_{\overline{b}\,\overline{b+1}}+\ldots + \Delta_{\overline{2n-1}\, \overline{2n}} + \Delta_{1\, \overline{2n}} + \ldots +\Delta_{a-1\,a}\,,\quad 1\leq a\leq b< 2n\,,
\end{aligned}
\end{equation}
and the $b = 2 n$ case of the latter is given by
\begin{equation}
    \Delta_{a\,\overline{2n}}=\Delta_{1\, \overline{2n}} + \ldots +\Delta_{a-1\,a}\,.
\end{equation}
One can easily verify that this construction is consistent with the definition \eqref{eq:BDeltarelation}
\begin{equation}
\begin{aligned}\label{eq:Delta_fromXY_consistency}
    \Delta_{ab}+\Delta_{a\bar{b}}=\Delta_{12}+\ldots +\Delta_{2n-1\,2n} + \Delta_{1\,\overline{2n}} = \sum_{i=1}^n\frac{X_i{+}Y_{i-1\,i}{+}Y_{i\,i+1}}{X}+\sum_{i=1}^n \frac{-2Y_{i\,i+1}}{X}=1\,,
\end{aligned}
\end{equation}
where $1\leq a\leq b\leq 2n$, and $\Delta_{a\,\bar{a}}=1$ is just a special case of the above. This map lets us express the symbol letters as
\begin{equation}
\begin{aligned}
    \frac{X^{+-}_{ij}}{X}=\Delta_{2i-1,2j+1}\,, && \frac{X^{++}_{ij}}{X}=\Delta_{2i-1,2j}\,,\\
    \frac{X^{--}_{ij}}{X}=\Delta_{2i-2,2j+1}\,, && \frac{X^{-+}_{ij}}{X}=\Delta_{2i-2,2j}\,,
\end{aligned}
\end{equation}
for $i\le j$; the remaining letters can be derived by using the relation \eqref{eq:BDeltarelation}.

\subsection{The Recursion for Wavefunction Coefficients}
\label{sec:recursion}
Our starting point for calculating $\psi^\circ_n$ is the differential equation for the wavefunction coefficient derived in \cite{He:2024olr}:
\begin{equation}\label{eq:dpsigon}
\begin{aligned}
    &\delta_{n-1,1} \psi_{n}^\circ (X;Y) 
    = \sum_{j=1}^n \left( \psi_{n-1}^\circ (\ldots,\tilde{X}_{j-1},\tilde{X}_{j+1},\ldots;\ldots,\tilde{Y}_{j-1\,j+1},\ldots)\otimes \log \frac{X_j^{+-}X_{j}^{-+}}{X_{j}^{--}X_j^{++}}\right.\\
     &\left.-\psi_{n-1}^\circ(\ldots,X_{j-1},X_{j}+X_{j+1},\ldots;\ldots,Y_{j{-}1\,j},Y_{j+1\,j+2},\ldots)\otimes\log \frac{X_j^{+-}X_{j+1}^{-+}}{X_{j}^{++}X_{j+1}^{++}}\right),
\end{aligned}
\end{equation}
where $\delta_{n-1,1}$ denotes the $(n-1,1)$ component of the coproduct\footnote{Note that $\delta$ is usually reserved for the cobracket, which is the coproduct modulo products; however to avoid clashes with our notation for the Pl\"ucker variables, we will use it here to indicate the full contribution to the coproduct, including product terms.}, and
\begin{align}
X_j^{\pm \pm} = X^{\pm\pm}_{jj}
\end{align}
with $X_{ij}^{\pm\pm}$ defined in \eqref{eq:xpmxpmdef}. In the first term of the sum the $j$-th site is deleted and its two neighbours are shifted
\begin{equation}\label{eq:tildeXY}
\begin{aligned}
\tilde{X}_{j-1} &= X_{j-1} + \tfrac{1}{2} X^{+-}_j\,,\\
\tilde{X}_{j+1} &= X_{j+1} + \tfrac{1}{2} X^{-+}_j\,,\\
\tilde{Y}_{j-1\,j+1} &= - \tfrac{1}{2} X^{--}_j\,,
\end{aligned}
\end{equation}
while in the second term the sites $j$ and $j{+}1$ are merged.
Therefore, \eqref{eq:dpsigon} expresses the coproduct of $\psi_{n}^{\circ}$ recursively in terms of $(n{-}1)$-gon wavefunction coefficients with shifted arguments~\eqref{eq:tildeXY}.

The parametrisation \eqref{eq:Delta_from_XY} significantly simplifies this recursion. Substituting it into the two final entries in \eqref{eq:dpsigon} and comparing with the cross-ratios \eqref{eq:normal_cross_ratios}, we find
\begin{equation}\label{eq:finalentries}
\begin{aligned}
    \log \frac{X_j^{+-}X_{j}^{-+}}{X_{j}^{--}X_j^{++}}
    &= \log\frac{\Delta_{2j-2,2j}\,\Delta_{2j-1,2j+1}}{\Delta_{2j-1,2j}\,\Delta_{2j-2,2j+1}}
    = \log\bigl(1-q_{2j-2,2j-1,2j,2j+1}\bigr)\,,\\[2pt]
    \log \frac{X_j^{+-}X_{j+1}^{-+}}{X_{j}^{++}X_{j+1}^{++}}
    &= \log\frac{\Delta_{2j-1,2j+1}\,\Delta_{2j,2j+2}}{\Delta_{2j-1,2j}\,\Delta_{2j+1,2j+2}}
    = \log\bigl(1-1/q_{2j-1,2j,2j+1,2j+2}\bigr)\,.
\end{aligned}
\end{equation}
Both final entries are therefore weight-one $A$-type quadrangular polylogarithms, as in~\cite{Ferro:2026oph}
\begin{equation}\label{eq:qliA1}
    \begin{aligned}
    \QLi^{+}_1(a,b,c,d) &= - \Li_1(q_{abcd})=\log(1-q_{abcd})\,,\\
    \QLi^{-}_1(a,b,c,d) &= +\Li_1(1/q_{abcd})=-\log(1-1/q_{abcd})\,,
\end{aligned}
\end{equation}
attached to the two quadrilaterals $(2j-2,2j-1,2j,2j+1)$ and $(2j-1,2j,2j+1,2j+2)$, which are the two quadrilaterals of the $4n$-gon corresponding to the deleted site. Note that the relative sign in \eqref{eq:dpsigon} is absorbed by the sign in the definition of $\text{QLi}_1^-$.

It remains to identify the shifted wavefunction coefficients. As in \cite{Ferro:2026oph}, the shifts \eqref{eq:tildeXY} act on the kinematics by deleting two adjacent marked points (and their conjugates) from the $4n$-gon. The first term of \eqref{eq:dpsigon} removes $z_{2j-1}$ and $z_{2j}$, while the second removes $z_{2j}$ and $z_{2j+1}$. Writing $\widehat{i}$ for the deleted $z_i$, the two sums combine into a single sum over the $2n$ adjacent pairs: 
\begin{equation}\label{eq:dgon}
\boxed{
    \delta_{n-1,1}\psi_n^\circ=\sum_{i=1}^{2n}\psi_{n-1}^\circ(\widehat{i},\widehat{i+1})\otimes\qli_1^{(-1)^{i+1}}(i-1,i,i+1,i+2)\,, \qquad n \ge 2\,,}
\end{equation}
with odd $i$ reproducing the first term of \eqref{eq:dpsigon} and even $i$ the second. We used the cyclic identification for $i=2n$: $\psi_{n-1}^\circ(\widehat{2n},\widehat{1}) = \psi^\circ_{n-1}(3,4,\cdots,2n-1,\overline{2})$.  We emphasise that \eqref{eq:dgon} is merely a rewriting of~(\ref{eq:dpsigon}); however, it provides a greatly simplified version of this recursion.  In the next subsection we solve the recursion explicitly.

%%%%%%%%%%%%%%%%%%%%%%%%%%%%%%%%%%%%%%%%%%%%%%%%%%%%%%%%%

\subsection{Solution of the Recursion}
\label{sec:solution}
Our main result is that the $n$-gon wavefunction coefficient in de Sitter space is given by
\begin{equation}\label{eq:Tarpppsi}
\boxed{
\begin{aligned}
 \psi_n^\circ
 &=\QLi_n^B
(1,\ldots,2n,\bar1,\ldots,\overline{2n})\,.
\end{aligned}}
\end{equation}
In contrast to the chain graph evaluated in \cite{Ferro:2026oph}, where the answer is a linear combination of three $F$-functions together with products of lower-weight functions, the entire $n$-gon wavefunction coefficient is a single weight-$n$ $B$-type quadrangular polylogarithm defined on the $4n$-gon. 

An alternative way to calculate $\psi_n^\circ$ is to use the cosmological tree theorem~\cite{AguiSalcedo:2023nds}, which expresses the $n$-gon wavefunction coefficient as a sum of terms, each of which is a (product of) chain graphs evaluated on specific kinematics. In contrast, \eqref{eq:Tarpppsi} takes a particularly simple form.

Here we outline the proof of \eqref{eq:Tarpppsi}. We will first show that the
$B$-type quadrangular polylogarithm function $\QLi^B_n$ defined in the previous section satisfies the same recursive coproduct formula as~\eqref{eq:dgon} for $n \ge 2$:
\begin{multline}
    \label{eq:QLiB_differential_eq}
    \delta_{n-1,1} \qli_{n}^B(1,\dots,2n,\bar{1},\dots,\overline{2n})=\\
    \sum_{i=1}^{2n} \qli_{n-1}^B(1,\dots,\widehat{i},\widehat{i{+}1},\dots,2n,\bar{1},\dots,\widehat{\overline{i}},\widehat{\overline{i{+}1}},\dots,\overline{2n})\otimes\qli_1^{(-1)^{i+1}}(i-1,i,i+1,i+2)\,.
\end{multline}
Here, only boundary quadrilaterals of the $4n$-gon contribute
to the coproduct. We use the general coproduct formula for
MPLs in~\eqref{eq:general-Li-coproduct}, which consists of weight-lowering terms and
two types of deletion of weight-one letters. From the
arborification rules~\eqref{eq:Aarbe}--\eqref{eq:tQB}, the contribution from
boundary quadrilaterals corresponds to the $i=1$ term in
the second sum of~\eqref{eq:general-Li-coproduct}. We therefore need to show that
all other terms cancel and that the surviving contribution
gives~\eqref{eq:QLiB_differential_eq}. The proof has two steps:
\begin{enumerate}
    \item In Appendix~\ref{sec:Annihilation_branches}, we first prove, by induction on the size of the branch, that the coproduct of the branches in~\eqref{eq:tQB} receives contributions only from boundary quadrilaterals. To do so, we rewrite the arborification rule in~\eqref{eq:Aarbe} so that the last weighted letter can correspond to a general alternating polygon rather than just a quadrilateral. We then group terms with the same weight left tensor factor and show that their logarithmic final entries cancel.
    \item Then in Appendix~\ref{sec:Annihilation_central_polygon}, we append the weighted letter associated with the self-conjugate central polygon in the $B$-type arborification and similarly show that the remaining logarithmic final entries sum to zero. 
\end{enumerate}
The surviving terms come from the boundary
quadrilaterals $E_i=(i-1,i,i+1,i+2)$. Removing $E_i$ and its
conjugate leaves precisely the lower-weight $B$-type
arborification. Applying $\Li$ then reproduces \eqref{eq:QLiB_differential_eq}.

Since the coproduct determines a function only up to additive constants of the appropriate weight, we finally rule out the possibility of any missing term by the same soft-limit argument used in \cite{Ferro:2026oph}. The wavefunction coefficient vanishes when any internal energy is taken soft~\cite{Arkani-Hamed:2023kig}, $Y_{m\, m+1}\propto\Delta_{2m\,2m+1}\to0$ for $m<n$ and $Y_{n\, 1}\propto \Delta_{1\,\overline{2n}}\to0$. The function \eqref{eq:Tarpppsi} satisfies this property, so there is no missing additive constant.

%%%%%%%%%%%%%%%%%%%%%%%%%%%%%%%%%%%%%%%%%%%%%%%%%%%%%%%%%
%%%%%%%%%%%%%%%%%%%%%%%%%%%%%%%%%%%%%%%%%%%%%%%%%%%%%%%%%

\section{Examples of \texorpdfstring{$n$}{n}-gon Wavefunction Coefficients in dS Cosmology}
\label{sec:examples}

%%%%%%%%%%%%%%%%%%%%%%%%%%%%%%%%%%%%%%%%%%%%%%%%%%%%%%%%%

In this section, we provide examples of symbols of $n$-gon wavefunction coefficients in de Sitter cosmology. These may be derived directly from our general formula in Section \ref{sec:physics}. Specifically we illustrate explicitly how the formula works for $n=1,2,3$.

\subsection{1-gon}

The 1-gon, or tadpole,
\begin{equation}
    \label{eq:tadpole} \psi_{1}^\circ(X_1,Y_{11})=\raisebox{-0.5cm}{\begin{tikzpicture}[vertex/.style={circle,fill=black,inner sep=1.6pt}]
  \def\r{0.6}
  \node[vertex,label=left:$X_1$] (v) at (0,0) {};
  \draw (\r,0) circle (\r);
  \node[right=3pt] at (2*\r,0) {$Y_{11}$};
\end{tikzpicture}}
\end{equation}
is a degenerate case: the $4n$-gon supports only the self-conjugate quadrilateral $(1,2,\bar{1},\bar{2})$, and therefore $\psi_1^\circ$ is the weight-one function built from the square-root letter $w_{12}$.
The map \eqref{eq:Delta_from_XY} gives variables
\begin{equation}
    \Delta_{12}=\frac{X_1+2Y_{11}}{X_1}\,,\qquad \Delta_{1\bar{2}}=1-\Delta_{12}=\frac{-2Y_{11}}{X_1}\,,
\end{equation}
with the latter being the single spurious letter. 

In Section \ref{sec:tarpBexample}, we calculated the weighted symbol at weight $n=1$ associated to the 4-gon \eqref{eq:BsquareB}. Further, we note that the symbol letter in \eqref{eq:BsquareB} is $w_{12}$, which is related to $\Delta_{12}$ by \eqref{eq:sqrtdefinition}. In Section \ref{subsubsec:Qliexamples}, we demonstrated that the Li map \eqref{eq:Li0R} applied to this weighted symbol gives \eqref{eq:QLiB1}:
\begin{align}
\label{eq:QLiB1new}
    \qli^B_1(1,2,\bar1,\bar2) = -\li_1(1/w_{12})=\log\left(\frac{1}{\Delta_{12}}\right) \,.
\end{align}
The case $n=1$ is the seed of the recursive differential equation, and we can check directly that this is the correct answer by evaluating the integral that defines this
wavefunction coefficient,
\begin{align}
\label{eq:oneloop_integral}
    \int_0^\infty d x_1 \frac{-2Y_{11}}{(X_1+2Y_{11}+x_1)(X_1+x_1)} = \log \left(\frac{X_1}{X_1+2Y_{11}}\right)=\log\left(\frac{1}{\Delta_{12}}\right).
\end{align}

%%%%%%%%%%%%%%%%%%%%%%%%%%%%%%%%%%%%%%%%%%%%%%%%%%%%%%%%%

\subsection{2-gon}
We turn to the 2-gon, or bubble graph,
\begin{equation}\label{eq:2gon_graph}
    \psi_{2}^\circ(X_1,Y_{12},X_2,Y_{21})=\raisebox{-1.2cm}{\begin{tikzpicture}[vertex/.style={circle,fill=black,inner sep=1.6pt}]
  \def\r{0.6}
  \draw (0,0) circle (\r);
  \node[vertex,label=left:$X_1$]  at (-\r,0) {};
  \node[vertex,label=right:$X_2$] at ( \r,0) {};
  \node[above=3pt] at (0, \r) {$Y_{12}$};
  \node[below=3pt] at (0,-\r) {$Y_{21}$};
\end{tikzpicture}}
\end{equation}
Here $n=2$, the algebra is $B_3$, and the relevant polygon is an octagon with vertices $\{1,2,3,4,\bar{1},\bar{2},\bar{3},\bar{4}\}$. 
The map~(\ref{eq:Delta_from_XY}) reads
\begin{equation}
\label{eq:bubblemap}
    \Delta_{12}=\frac{X_1+Y_{12}+Y_{21}}{X}\,,\qquad
    \Delta_{23}=\frac{-2Y_{12}}{X}\,,\qquad
    \Delta_{34}=\frac{X_2+Y_{12}+Y_{21}}{X}\,,
\end{equation}
and generates the dictionary
\begin{equation}
\label{eq:bubbledict}
\begin{aligned}
    \Delta_{13}&=\frac{X_1-Y_{12}+Y_{21}}{X}\,, &\quad
    \Delta_{24}&=\frac{X_2-Y_{12}+Y_{21}}{X}\,, &\quad
    \Delta_{14}&=\frac{X_1+X_2+2Y_{21}}{X}\,,\\
    \Delta_{1\bar 2}&=\frac{X_2-Y_{12}-Y_{21}}{X}\,, &\quad
    \Delta_{3\bar 4}&=\frac{X_1-Y_{12}-Y_{21}}{X}\,, &\quad
    \Delta_{2\bar 3}&=\frac{X_1+X_2+2Y_{12}}{X}\,,\\
    \Delta_{1\bar 3}&=\frac{X_2+Y_{12}-Y_{21}}{X}\,, &\quad
    \Delta_{2\bar 4}&=\frac{X_1+Y_{12}-Y_{21}}{X}\,, &\quad
    \Delta_{1\bar 4}&=\frac{-2Y_{21}}{X}\,,
\end{aligned}
\end{equation}
which reproduces the ten letters in the symbol $\mathcal{S}(\psi_2^\circ)$ together with the two
spurious letters $\Delta_{23}$ and $\Delta_{1\bar 4}$.

The 2-gon wavefunction coefficient can then be determined recursively from $\psi_1^\circ$ using \eqref{eq:dgon}:
\begin{equation}
\begin{aligned}
    \delta_{1,1}\psi_2^\circ= &\sum_{i=1}^{4}\psi_{1}^\circ(\widehat{i},\widehat{i+1})\otimes\qli_1^{(-1)^{i+1}}(i-1,i,i+1,i+2)\\
    = &- \psi_1^\circ(3,4)\otimes \li_1(q_{\bar 4123})+ \psi_1^\circ(1,4)\otimes \li_1(1/q_{1234})\\
    &- \psi_1^\circ(1,2)\otimes \li_1(q_{234\bar1})+ \psi_1^\circ(3,\bar 2)\otimes \li_1(1/q_{34\bar1\bar2})\,.
\end{aligned}
\end{equation}
Here $\psi_1^\circ$ is given by~(\ref{eq:QLiB1new}), $\QLi_1^\pm$ by~(\ref{eq:qliA1}), and the cross-ratios by~(\ref{eq:normal_cross_ratios}) and~(\ref{eq:sqrtdefinition}), using the coordinate map~(\ref{eq:bubblemap})--(\ref{eq:bubbledict}).
The solution to the above recursion is
\begin{align}
\label{eq:2gonQli}
    \psi_2^\circ= \qli^B_2(1,2,3,4,\bar1,\bar2,\bar3,\bar4) \,,
\end{align}
where $\qli_2^B$ is as defined in \eqref{eq:BQLi2}. In principle, there could have been an additive constant but the soft limit $Y_{12}\to0$ confirms that \eqref{eq:2gonQli} is exact. This matches the result for the 2-gon in \cite{Hillman:2019wgh} as well as a numerical evaluation of the associated integral.

The structure of the recursion relation is also mirrored by that of the symbol written in terms of cross-ratios:
\begin{equation}
\begin{aligned}
\label{eq:psi2symbol}
    \mathcal{S}(\psi_2^\circ)&=(1-1/w_{12})\otimes (1-q_{234\bar1})-(1-1/w_{14})\otimes (1-q_{2341}) \\
    &+(1-1/w_{34})\otimes (1-q_{\bar2\bar34\bar1}) -(1-w_{23})\otimes (1-q_{\bar2 34 \bar1})\\
    &=\mathcal{S}\left[\psi_1^\circ(1,2)\right]\otimes (1-q_{234\bar1})- \mathcal{S}\left[\psi_1^\circ(1,4)\right]\otimes (1-q_{2341}) \\
    &+\mathcal{S}\left[\psi_1^\circ(3,4)\right]\otimes (1-q_{\bar2\bar34\bar1}) -\mathcal{S}\left[\psi_1^\circ(3,\bar{2})\right]\otimes (1-q_{\bar2 34 \bar1})\,.
\end{aligned}
\end{equation}
One can easily check that this expression matches the symbol of $\qli^B_2(1,2,3,4,\bar1,\bar2,\bar3,\bar4)$ as defined in \eqref{eq:BQLi2}.

%%%%%%%%%%%%%%%%%%%%%%%%%%%%%%%%%%%%%%%%%%%%%%%%%%%%%%%%%

\subsection{3-gon}
The same steps may be carried out for the wavefunction coefficient of the 3-gon, or triangle graph,
\begin{equation}
    \psi_{3}^\circ(X_1,Y_{12},X_2,Y_{23},X_3,Y_{31})=\raisebox{-1.2cm}{\begin{tikzpicture}[vertex/.style={circle,fill=black,inner sep=1.6pt}]
  \def\r{0.8}
  \draw (0,0) circle (\r);
  \node[vertex,label=left:$X_1$]        at (180:\r) {};
  \node[vertex,label=above right:$X_2$] at ( 60:\r) {};
  \node[vertex,label=below right:$X_3$] at (-60:\r) {};
  \node at (120:{\r+0.4}) {$Y_{12}$};
  \node at (  0:{\r+0.5}) {$Y_{23}$};
  \node at (240:{\r+0.3}) {$Y_{31}$};
\end{tikzpicture}}
\end{equation}
whose symbol has weight three. Here $n=3$, the algebra is $B_5$, and the polygon is a dodecagon with vertices $\{1,\ldots,6,\bar{1},\ldots,\bar{6}\}$. The map \eqref{eq:Delta_from_XY} gives
\begin{equation}
\begin{aligned}
    \label{eq:Delta3gon}& \Delta_{12}=\frac{X_1+Y_{12}+Y_{31}}{X}\,,\qquad \Delta_{23}=\frac{-2Y_{12}}{X}\,,\qquad \Delta_{34}=\frac{X_2+Y_{12}+Y_{23}}{X}\,,&\\
    &\Delta_{45}=\frac{-2Y_{23}}{X}\,,\qquad \Delta_{56}=\frac{X_3+Y_{23}+Y_{31}}{X}\,,\qquad \Delta_{1\bar{6}}=\frac{-2Y_{31}}{X}&
\end{aligned}
\end{equation}
with $X=X_1+X_2+X_3$.

The wavefunction coefficient $\psi_3^\circ$ can be derived from $\psi_2^\circ$ via the recursion relation \eqref{eq:dgon}:
\begin{equation}
\begin{aligned}
\label{eq:dgon3}
    \delta_{2,1}\psi_3^\circ =  &\sum_{i=1}^{6}\psi_{2}^\circ(\widehat{i},\widehat{i+1})\otimes\qli_1^{(-1)^{i+1}}(i-1,i,i+1,i+2)\\
    =&- \psi_2^\circ(3,4,5,6)\otimes \li_1(q_{\bar 6123})+ \psi_2^\circ(1,4,5,6)\otimes \li_1(1/q_{1234})\\
    &- \psi_2^\circ(1,2,5,6)\otimes \li_1(q_{2345})+ \psi_2^\circ(1,2,3,6)\otimes \li_1(1/q_{3456})\\
    &- \psi_2^\circ(1,2,3,4)\otimes \li_1(q_{456\bar{1}})+ \psi_2^\circ(3,4,5,\bar2)\otimes \li_1(1/q_{56\bar1\bar2})\,,
\end{aligned}
\end{equation}
where $\psi^\circ_2$ is as given in \eqref{eq:2gonQli} and the cross-ratios can be determined using \eqref{eq:normal_cross_ratios}, \eqref{eq:sqrtdefinition} and the $\Delta$ map \eqref{eq:Delta3gon}.

Once again we can determine the solution to this recursion (using the soft limit $Y_{12}, Y_{23}, Y_{31}\to0$ to fix the additive constant to 0) to be
\begin{align}
    \psi^\circ_3 = \qli^B_3(1,2,3,4,5,6,\bar1,\bar2,\bar3,\bar4,\bar5,\bar6)\,,
\end{align}
where $\qli^B_3$ is as defined in \eqref{eq:QliB3}. As far as we know, no explicit result for this wavefunction coefficient has appeared before in the literature, but we have successfully compared it against explicit numerical evaluations  of the integral representation for $\psi^\circ_3$.

The symbol of $\psi_3^\circ$ can be written in a compact form that manifests the recursive structure \eqref{eq:dgon3},
\begin{equation}
\begin{aligned}
\label{eq:psi3symbol}
\mathcal{S}(\psi_3^\circ)&= \mathcal{S}(\psi_2^\circ(3,4,5,6))\otimes (1-q_{\bar 6123})-\mathcal{S}(\psi_2^\circ(1,4,5,6))\otimes (1-q_{2341})\\
&+\mathcal{S}(\psi_2^\circ(1,2,5,6)) \otimes (1-q_{2345})-\mathcal{S}(\psi_2^\circ(1,2,3,6)) \otimes (1-q_{4563})\\
&+\mathcal{S}(\psi_2^\circ(1,2,3,4)) \otimes (1-q_{456\bar{1}}) -\mathcal{S}(\psi_2^\circ(3,4,5,\bar2))\otimes (1-q_{6\bar{1}\bar{2}5})\,,
\end{aligned}
\end{equation}
where $\mathcal{S}(\psi^\circ_2)$ is as given in \eqref{eq:psi2symbol}. One cross-check is that the symbol above matches that of $\qli^B_3(1,2,3,4,5,6,\bar1,\bar2,\bar3,\bar4,\bar5,\bar6)$ as given in \eqref{eq:QliB3}.

%%%%%%%%%%%%%%%%%%%%%%%%%%%%%%%%%%%%%%%%%%%%%%%%%%%%%%%%%
%%%%%%%%%%%%%%%%%%%%%%%%%%%%%%%%%%%%%%%%%%%%%%%%%%%%%%%%%

\section{Chain Graphs as a Limit of Polygon Graphs}
\label{sec:limit2}
From the point of view of Feynman graphs, the $n$-gon and the $n$-site chain are related to each other by removing a single internal line of the $n$-gon. In this section we show that this operation can be understood at the level of the wavefunction coefficients associated to these graphs, and demonstrate that there exists a limit of the $B$-type quadrangular polylogarithm  $\QLi^B_n = \psi^\circ_n$ that exactly returns the combination of $A$-type quadrangular polylogarithms present in the wavefunction coefficient $\psi_n$ of the chain graph.
We first consider the effect of this kinematic limit on the integrand before noting that it immediately lifts to a function-level statement. We follow this with examples to illustrate how the limit operates on symbols and functions. 

\subsection{Kinematic Limit}
The wavefunction coefficient of the $n$-gon graph in de Sitter space is given by
\begin{align}
\label{eq:flat_integral}
    \psi_n^\circ (\vec X,\vec Y)= \int_0^\infty \prod_{i=1}^n d x_i\, \psi_n^{\circ,\,\text{flat space}}(\vec X+\vec x,\vec Y)\,,
\end{align}
where $\psi_n^{\circ,\,\text{flat space}}$ is the wavefunction coefficient in flat space. This can be calculated by considering maximal non-intersecting tubings of a graph giving rise to the formula
\begin{equation}
\label{eq:integrandXY}
\begin{aligned}
    &\psi_n^{\circ,\,\text{flat space}}(\vec X,\vec Y) = \sum_{j=1}^n \hat \psi_n^{\text{flat space}}(\vec X, \vec Y; Y_{j\,j+1})\times\frac{-2Y_{j\,j{+}1}}{X}\\
    &\hat \psi_n^{\text{flat space}}(\vec X, \vec Y; Y_{j\,j+1}) = 
    \begin{tikzpicture}[baseline=(current bounding box.center),
    dot/.style={circle, fill=black, inner sep=1.6pt}]
    % vertices
    \node[dot, label={[label distance=0pt]below:$X_{j+1}+Y_{j\,j+1}$}] (x1) at (0,0) {};
    \node[dot, label={[label distance=0pt]below:$X_{j+2}$}] (x2) at (2.5,0) {};
    \node[dot, label={[label distance=0pt]below:$X_{j-1}$}] (x3) at (5,0) {};
    \node[dot, label={[label distance=0pt]below:$X_j+Y_{j\,j+1}$}] (x4) at (7.5,0) {};
    % edges
    \draw[thick] (x1) -- node[above=1pt] {$Y_{j+1\,j+2}$} (x2);
    \draw[thick] (x3) -- node[above=1pt] {$Y_{j-1\,j}$} (x4);
    % connecting stub lines and dots
    \draw[thick] (x2) -- (3.4,0);
    \node at (3.75,0) {$\cdots$};
    \draw[thick] (4.1,0) -- (x3);
    \end{tikzpicture}
\end{aligned}
\end{equation}
where $\hat \psi_n^{\text{flat space}}$ is the flat space $n$-site chain wavefunction coefficient with shifted kinematics at the two endpoints.
We claim that this $n$-gon integrand degenerates to the $n$-site chain integrand under the following two operations:
\begin{equation}
\begin{aligned}\label{eq:polygon_to_chain_operation}
    &\text{1. A shift: }X_1 \to X_1 - Y_{n\,1},\quad X_n \to X_n - Y_{n\,1},\\
    &\text{2. A limit: }Y_{n\,1}\to\infty.
\end{aligned}
\end{equation}

We can see that the shift operation in \eqref{eq:polygon_to_chain_operation} undoes the endpoint kinematic shifts of the $j=n$ term in \eqref{eq:integrandXY}: $\hat \psi_n^{\text{flat space}}(\vec X, \vec Y; Y_{n\,1})\to \psi_n^{\text{flat space}}(\vec X, \vec Y)$. In addition, the ratio in the summand is shifted to $-2Y_{j\,j+1}/(X)\to -2Y_{j\,j+1}/(X-2Y_{n\,1})$. After taking the limit, all of the ratios in the summand vanish except in the $j=n$ term: $-2Y_{n\,1}/(X-2Y_{n\,1})\to 1$. This proves the claim that the $n$-gon integrand reduces to the $n$-site chain integrand under the operation \eqref{eq:polygon_to_chain_operation}.

Now we proceed from the integrand to the integral $\psi_n^\circ$. The case $n=1$ is special, because taking the shift $X_1\to X_1 - 2Y_{11}$, followed by the limit $Y_{11}\to\infty$ causes the integral to diverge. As a result, we \emph{cannot} extend our analysis from the 1-gon integrand to the 1-gon integral. This divergence can be understood via a method of regions analysis in the $x_1$ integration variable in \eqref{eq:oneloop_integral}. In particular, although the integrand in \eqref{eq:oneloop_integral} vanishes when $0\ll Y_{11}\ll x_1$, it diverges in the region $0\ll Y_{11}\sim x_1$. A similar analysis for $n\ge 2$ shows that there are enough factors of $x_i$ in the denominator of the integrand to prevent any such divergence.   This argument establishes that\footnote{Comparing numerical values of the $\qli^B_n$ formula for the $n$-gon presented in this paper with the $\qli$ formula for the $n$-site chain in \cite{Ferro:2026oph}, we find that the two formulas differ by an overall sign under the operation \eqref{eq:polygon_to_chain_operation}. This is a difference in conventions; the de Sitter integrand used in \cite{Ferro:2026oph} is the negative of the flat space wavefunction.}
\begin{equation}
\label{eq:shiftlimit}
\boxed{\psi_n^\circ (\vec X,\vec Y) \overset{\text{shift}}{\underset{\text{limit}}{-\hspace{-2mm}-\hspace{-2mm}-\hspace{-2mm}\longrightarrow}} \psi_n (\vec X,\vec Y)\qquad\text{for }n\ge 2\,.}
\end{equation}

\subsection{Limits of the Cluster Variables and Symbols}
\label{sec:sixpointtwo}

We can translate the kinematic limit discussed above into $\Delta$ variables using the definitions presented in Section \ref{subsec:kinematics}. Let us start by taking the limit of the cluster variables associated to the edges of the auxiliary $4n$-gon (as defined in \eqref{eq:Delta_from_XY}),
\begin{equation}
\label{eq:delta_limit}
\begin{aligned}
    \Delta_{1\,2} \to \frac{X_1+Y_{1\,2}}{-2Y_{n\,1}}\,,&\quad\Delta_{2a-1\,2a} \to \frac{X_a+Y_{a-1\,a}+Y_{a\,a+1}}{-2Y_{n\,1}}\,, &&1<a<n\,,\\
    \Delta_{2n-1\,2n} \to \frac{X_n+Y_{n-1\,n}}{-2Y_{n\,1}}\,,&\quad\Delta_{2a\,2a+1} \to \frac{Y_{a\,a+1}}{Y_{n\,1}}\,,&&1\le a<n\,,\\
    \Delta_{1\,\overline{2n}} \to 1\,.
\end{aligned}
\end{equation}
The other $\Delta_{ab}$ are then given by \eqref{eq:Delta_fromXY2}. This matches the map from kinematic variables $X_i$ and $Y_{i\,i{+}1}$ to $A$-type cluster variables first presented in \cite{Capuano:2025myy}.

In fact, there is a natural way to understand this limit in terms of the polygonal construction of $A$- and $B$-type cluster algebras, as explained in Section \ref{sec:review}. Importantly, the natural setting for $B$-type is the folded description we explain in Appendix \ref{app:Bcoords}, in which the $B_{2n-1}$ algebra is realised on a once-punctured $2n$-gon rather than on the centrally symmetric $\mathbf{P}_{4n}$. The limit is then the degeneration in which the puncture moves to the boundary of the $2n$-gon, and the once-punctured $2n$-gon becomes an unpunctured $(2n+1)$-gon, precisely the setting on which the chain functions of \cite{Ferro:2026oph} are defined, and therefore $B_{2n-1}\to A_{2n-2}$. For this degeneration, the rank drops by one, and the $\mathbb{Z}_2$ involution disappears.

It is a straightforward exercise to check for the first few $n$ that the symbol of the $n$-gon wavefunction coefficient $\mathcal{S}(\psi^\circ_n)$ written in terms of the cluster variables $\Delta_{ij}$ reduces to the symbol of the $n$-site chain wavefunction coefficient $\mathcal{S}(\psi_n)$ in the limit described above.

Here we illustrate the example $n=2$.
Upon shifting and taking the limit, the 12 cluster variables of $B_3$ become
\begin{equation}
\begin{aligned}
    &\Delta_{12}\to\frac{X_1+Y_{12}}{-2Y_{21}}\,,\quad& \Delta_{23}&\to\frac{Y_{12}}{Y_{21}}\,,\quad &\Delta_{34}&\to\frac{X_2+Y_{12}}{-2Y_{21}}\,,\\
    &\Delta_{14}\to\frac{X}{-2 Y_{21}}\,,\quad& 
    \Delta_{13}&\to\frac{X_1-Y_{12}}{-2Y_{21}}\,,\quad& \Delta_{24}&\to\frac{X_2-Y_{12}}{-2Y_{21}}\,,\\
    &\Delta_{1\bar2}\to1\,,\quad& \Delta_{2\bar3}&\to1\,,\quad&\Delta_{3\bar4}&\to1\,,\\
    &\Delta_{1\bar4}\to1\,,\quad&
    \Delta_{1\bar3}&\to1\,,\quad& \Delta_{2\bar4}&\to1\,.
\end{aligned}
\end{equation}
There are 6 nontrivial cluster variables left after this operation, from which we can form five ratios that remain finite in the limit $Y_{21} \to \infty$; these are the 5 cluster variables of $A_2$.

The symbol of the 2-gon wavefunction coefficient reads
\begin{align}\label{eq:Spsi2circ}
       \mathcal{S}(\psi_{2}^\circ)= \Delta_{12}\otimes \frac{\Delta_{34} \Delta_{1\bar{2}}}{\Delta_{24} \Delta_{1\bar{3}}}+\Delta _{2\bar{3}}\otimes \frac{\Delta_{1\bar{3}} \Delta_{2\bar{4}}}{\Delta_{12} \Delta_{34}}+\Delta_{34}\otimes \frac{\Delta_{12} \Delta_{3\bar{4}}}{\Delta_{13} \Delta_{2\bar{4}}}+\Delta_{14}\otimes \frac{\Delta_{13} \Delta_{24}}{\Delta_{12} \Delta_{34}}\,.
\end{align}
In the limit \eqref{eq:delta_limit}, the variables $X_{ij}^{\pm\pm}$ that survive are exactly those of the extended chain graph, and therefore \eqref{eq:Spsi2circ} can be rewritten back in the $A$-type variables of \cite{Capuano:2025myy,Ferro:2026oph} as
\begin{align}
\label{eq:Spsi2_chain}
       \mathcal{S}(\psi_{2}^\circ)\to \Delta_{12}\otimes \frac{\Delta_{34} }{\Delta_{24}}+\Delta_{34}\otimes \frac{\Delta_{12} }{\Delta_{13}}+\Delta_{14}\otimes \frac{\Delta_{13} \Delta_{24}}{\Delta_{12} \Delta_{34}}= \frac{\Delta_{12}}{\Delta_{14}}\otimes \frac{\Delta_{34} }{\Delta_{24}}+\frac{\Delta_{34}}{\Delta_{14}}\otimes \frac{\Delta_{12} }{\Delta_{13}}
\end{align}
which is $\mathcal{S}(\psi_2)$, eq.~(2.4) of \cite{Ferro:2026oph}. Note that the five letters appearing in \eqref{eq:Spsi2_chain} are cluster variables of $A_2$, as anticipated above. We have explicitly verified by direct computation that~(\ref{eq:shiftlimit}) holds at symbol level through $n=4$.

\subsection{Limits of the Functions}

The same statement holds at the level of the functions, and in particular relates transcendental terms which the symbol cannot see. Once again, we illustrate this via the 2-gon example, in which case the wavefunction coefficient can be written in terms of multiple polylogarithms as
\begin{equation}
\begin{aligned}
    \psi^\circ_2
    &= \qli^B_2(1,2,3,4,\bar1,\bar2,\bar3,\bar4)\\
    &=\text{Li}_{1,1} \left(\frac{\Delta_{23}\Delta_{4\bar 1}}{\Delta_{34}\Delta_{\bar 12}},-\frac{\Delta_{\bar 12}}{\Delta_{12}}\right) -\text{Li}_{1,1}\left(\frac{\Delta_{4\bar 1}\Delta_{\bar 23}}{\Delta_{34}\Delta_{12}},-\frac{\Delta_{23}}{\Delta_{\bar 23}}\right)
    + \text{Li}_{1,1}\left(\frac{\Delta_{4\bar 1}\Delta_{23}}{\Delta_{12}\Delta_{\bar 34}},-\frac{\Delta_{\bar 34}}{\Delta_{34}}\right)\\ &- \text{Li}_{1,1}\left(\frac{\Delta_{23}\Delta_{41}}{\Delta_{12}\Delta_{34}},-\frac{\Delta_{\bar 14}}{\Delta_{14}}\right)
     +\text{Li}_{2}\left(-\frac{\Delta_{23}\Delta_{4\bar 1}}{\Delta_{12}\Delta_{34}}\right),
\end{aligned}
\end{equation}
where we have used the definition~\eqref{eq:BQLi2}. To take the limit, we need the asymptotic behaviour of $\text{Li}_{1,1}$ and $\text{Li}_2$ for large arguments,
\begin{align}
       \lim_{y\to\infty} \text{Li}_{1,1}(x,y)&=\tfrac{1}{2}\log^2(x(1-y))+\text{Li}_2(1-x)+\mathcal{O}\left(\frac{1}{y}\right),\\
              \lim_{x\to\infty} \text{Li}_{2}(x)&=-\tfrac{1}{2}\log^2(-x)-\frac{\pi^2}{6}+\mathcal{O}\left(\frac{1}{x}\right),\\
 \lim_{\substack{x,y\to\infty\\ x/y\ \mathrm{fixed}}} \text{Li}_{1,1}(x y,1/y)&=\mathcal{O}\left(\frac{1}{y}\right).
\end{align}
All four $\log^2$ contributions cancel against one another, leaving the combination
\begin{align}
\psi_2^\circ&\to\text{Li}_2\left(\frac{\Delta_{24}}{\Delta_{34}}\right)+\text{Li}_2\left(\frac{\Delta_{13}}{\Delta_{12}}\right)-\text{Li}_2\left(\frac{\Delta_{13}\Delta_{24}}{\Delta_{12}\Delta_{34}}\right)-\frac{\pi^2}{6}\,,
\end{align}
which is the correct expression for $\psi_2$, including the $\pi^2$ term. We have explicitly checked that~(\ref{eq:shiftlimit}) holds at function level up to $n=4$.

%%%%%%%%%%%%%%%%%%%%%%%%%%%%%%%%%%%%%%%%%%%%%%%%%%%%%%%%%
%%%%%%%%%%%%%%%%%%%%%%%%%%%%%%%%%%%%%%%%%%%%%%%%%%%%%%%%%

\section{Conclusions}
In this paper we have shown that the de Sitter space cosmological wavefunction coefficients $\psi_n^\circ$ associated to  $n$-gon graphs (before integrating over the loop momentum) are given by a new class of transcendental functions, which we call $B$-type quadrangular polylogarithms. We have constructed these functions explicitly by extending the work of Matveiakin and Rudenko to the $B$-type cluster algebras that control the symbol alphabets of the $n$-gon wavefunction coefficients, using an equivariant version of the arborification procedure adapted to quadrangulations of a $4n$-gon invariant under a $180^\circ$ rotation. A key feature of this construction is that each centrally symmetric partial quadrangulation has a unique self-conjugate $4k$-gon, which fixes a canonical root for the dual tree. For $k=1$, the weighted symbol letter is built from a square root of the self-conjugate cross-ratio rather than the cross-ratio itself; this last feature makes the $B$-type polylogarithms a genuinely new family of functions, not simply related to the $A$-type ones.

Using these functions, the wavefunction coefficient $\psi_n^\circ$ of the $n$-gon graph, for $n \geq 1$, takes a remarkably compact form: it is a single weight-$n$ $B$-type quadrangular polylogarithm $\qli^B_n$, with no additional products of lower-weight pieces or additive constants. We verified this explicitly for the one\nobreakdash-, two-, three-, and four-gon graphs, corresponding respectively to quadrangulations of the square, octagon, dodecagon, and hexadecagon, and proved it in general by showing that our closed-form expression satisfies the differential recursion relating loop-level wavefunction coefficients to their lower-point counterparts.

The key property of $\psi_n^\circ=\qli^B_n$ is that its symbol alphabet consists of cluster variables of the $B_{2n-1}$ cluster algebra, and its symbol satisfies total compatibility.
Remarkably, we also show that for $n \ge 2$, in  a particular kinematic limit, the $B$-type quadrangular polylogarithm $\qli^B_n$ exactly degenerates to the combination of $A$-type quadrangular polylogarithms that compute the wavefunction coefficient $\psi_n$ of the $n$-site chain graph.

It is natural to ask whether these constructions extend beyond the one-loop graphs considered here to multi-loop diagrams or even more general graphs where $n$-gons appear as subgraphs of a larger graph. It is clear that the symbol of any wavefunction coefficient in de Sitter space satisfies total compatibility with respect to its symbol entries, which however are cluster variables only in the special cases of chain or polygon graphs, considered respectively in~\cite{Ferro:2026oph} and this paper. 
Another interesting open problem is to integrate our formulas, say for example $\psi_3^\circ$, over the loop momentum; an integrated result for the triangle graph in $\phi^4$ theory has recently been found in~\cite{Henn:2026lfz}.
Finally, it would be interesting to study whether the cluster-algebraic structure survives at the level of the cosmological correlators.

On the mathematical side, let us recall that the construction of Matveiakin and Rudenko gives an infinite sequence of polylogarithm functions of arbitrary weight $\ge n-1$ that satisfy total compatibility with respect to the $A_{2n-2}$ cluster algebra, although only the two lowest-weight members of this sequence play a role in the chain graph wavefunction coefficients. The polygon graphs considered here are even simpler, in that we only needed the single weight-$n$ $\qli_n^B$ function; presumably this is also the lowest-weight member in an infinite sequence and it would be interesting to construct its higher-weight siblings, and to fully classify/enumerate the weight $w$ functions satisfying total compatibility with respect to $B_m$ for arbitrary $w$ and $m$.
We leave all of these questions to future work.

%%%%%%%%%%%%%%%%%%%%%%%%%%%%%%%%%%%%%%%%%%%%%%%%%%%%%%%%%
%%%%%%%%%%%%%%%%%%%%%%%%%%%%%%%%%%%%%%%%%%%%%%%%%%%%%%%%%

\section*{Acknowledgements}
This work was supported in part by the Deutsche Forschungsgemeinschaft (DFG, German Research Foundation) Projektnummer 508889767/FOR5582 ``Modern Foundations of Scattering Amplitudes" (LF, Y-QZ), by the Royal Society via a Newton International Fellowship (LR), by the US Department of Energy under contract DE-SC0010010 Task F (SP, MS, AV, H-CW), and by Simons Investigator Award \#376208 (SP, AV, H-CW). This research was also supported by the Munich Institute for Astro-, Particle and BioPhysics (MIAPbP) which is funded by the Deutsche Forschungsgemeinschaft (DFG, German Research Foundation) under Germany's Excellence Strategy -- EXC-2094 -- 390783311. AI systems were used to generate figures, draft the first version of the proof presented in Appendix C, proofread, and generate Mathematica files to audit the calculations. All resulting material was reviewed and verified by the authors.

%%%%%%%%%%%%%%%%%%%%%%%%%%%%%%%%%%%%%%%%%%%%%%%%%%%%%%%%%
%%%%%%%%%%%%%%%%%%%%%%%%%%%%%%%%%%%%%%%%%%%%%%%%%%%%%%%%%
%%%%%%%%%%%%%%%%%%%%%%%%%%%%%%%%%%%%%%%%%%%%%%%%%%%%%%%%%

\appendix 

\section{Equivalent Descriptions of \texorpdfstring{$B$}{B}-Type Cluster Algebras}

\label{app:Bcoords}
In this appendix we discuss the equivalence of two geometric descriptions of the cluster coordinates of $B_m$ mentioned in Section \ref{sec:review}, namely:
\begin{enumerate}
    \item the folding of a once-punctured $(m+1)$-gon and its tagged triangulations, and
    \item the $(2m+2)$-gon $\mathbf{P}_{2m+2}$ and its centrally symmetric triangulations.
\end{enumerate}

The first description comes from folding the $D_{m+1}$ cluster algebra  by the $\mathbb{Z}_2$ automorphism exchanging the two nodes of the fork in the Dynkin diagram, as outlined in \cite{fomin2021introductionclusteralgebraschapters}. We start by taking an $(m+1)$-gon with a puncture $P$ and labelling its boundary vertices $1,\ldots,m+1$ clockwise. An arc joining two boundary vertices $i\ne j$ may pass on either side of $P$, and therefore each unordered pair supports two distinct arcs. We write $\tilde{\Delta}_{ij}$ for the one that runs from $i$ to $j$ clockwise around $P$, so that $\tilde\Delta_{ij}$ and $\tilde\Delta_{ji}$ are the two arcs of the pair. They satisfy
\begin{align}
    \tilde\Delta_{ij}=1-\tilde\Delta_{ji}\,, \qquad\text{for }i>j\,,
\end{align}
which is the punctured-polygon equivalent of the relation $\Delta_{ab}+\Delta_{a\bar{b}}=1$. In addition, there are the arcs $\tilde\Delta_{iP}$ ending on the puncture itself. In $D_{m+1}$ each of these carries two tags, and folding identifies the two tagged copies into a single variable $\tilde\Delta_{iP}$.

Arcs are compatible if and only if they do not cross, with one exception: two distinct arcs $\tilde\Delta_{iP}$ and $\tilde\Delta_{jP}$ are considered incompatible, even though they meet only at $P$. This exception can be understood naturally in the unfolded picture: the arcs to the puncture lift to the long diagonals of $\mathbf{P}_{2m+2}$, and any two of those cross at the centre. An arc to the puncture is compatible with a boundary-to-boundary arc precisely when the two do not intersect in the interior, in exactly the same way as for any other pair.

To illustrate this, consider $B_3$, obtained by folding $D_4$ on a punctured square:
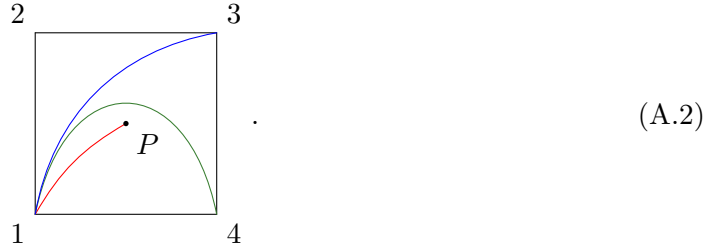
\begin{align}
    \begin{tikzpicture}[scale=0.4,line cap=round,line join=round, arr/.style={->, >=stealth},baseline=(current bounding box.center)]
    % triangle vertices
    \coordinate (BL) at (-3,0);
    \coordinate (BR) at (3,0);
    \coordinate (TL) at (-3,6);
    \coordinate (TR) at (3,6);
    \node[below left]  at (BL) {$1$};
    \node[below right] at (BR) {$4$};
    \node[above left]  at (TL) {$2$};
    \node[above right] at (TR) {$3$};
    % puncture
    \coordinate (P) at (0,3);
    \node[below right]  at (P) {$P$};
%    \draw[green] (P) -- (-3,3); 
    % boundary triangle
    \draw (BL) --  (BR);
    \draw (BR) --  (TR);
    \draw (TR) --  (TL);
    \draw (TL) --  (BL);
    % two arcs from left vertex to puncture
    \draw[red] (BL) to[bend left = 15]  (P);
    %\draw[red] (BL) to[bend right = 15]  (P);
    % big arc from BL to BR
    \draw[OliveGreen] (BL) .. controls (-2.0,4.9) and (2.0,4.9) ..  (BR);
    % big arc from BL to TR: outermost arc
    \draw[blue] (BL) to[out=80,in=190]  (TR);
    % puncture
    \fill (P) circle (2.5pt);
    \end{tikzpicture}.
\end{align}
The picture contains a triangulation of the once-punctured square consisting of chords $\textcolor{blue}{\tilde\Delta_{13}}$, $\textcolor{OliveGreen}{\tilde\Delta_{14}}$ and $\textcolor{red}{\tilde\Delta_{1P}}$. Note that the red chord $\tilde\Delta_{1P}$ begins as two radii in the original $D$-type algebra and is folded into one chord $\tilde\Delta_{1P}$ in the geometric description of $B$-type coordinates.

The second description is the one used in Section \ref{sec:review}. For B$_3$, it involves an octagon with vertices labelled $1,\ldots,4,\bar{1},\ldots,\bar{4}$ in clockwise cyclic order:
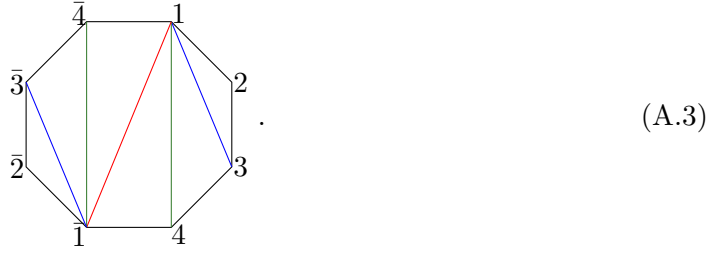
\begin{align}
    \begin{tikzpicture}[
    scale=0.4,
    circ/.style={draw,circle,fill=white,minimum size=8pt,inner sep=0pt},
    arr/.style={->},baseline=(current bounding box.center)
    ]
    % octagon vertices
    \coordinate (v1) at ( 1.4,  3.4);
    \coordinate (v2) at ( 3.4,  1.4);
    \coordinate (v3) at ( 3.4, -1.4);
    \coordinate (v4) at ( 1.4, -3.4);
    \coordinate (v5) at (-1.4, -3.4);
    \coordinate (v6) at (-3.4, -1.4);
    \coordinate (v7) at (-3.4,  1.4);
    \coordinate (v8) at (-1.4,  3.4);
    % boundary
    \draw (v1)--(v2)--(v3)--(v4)--(v5)--(v6)--(v7)--(v8)--cycle;
    % vertex labels
    \node at ($(v1)+(0.25,0.25)$) {$1$};
    \node at ($(v2)+(0.3,0)$) {$2$};
    \node at ($(v3)+(0.3,0)$) {$3$};
    \node at ($(v4)+(0.25,-0.25)$) {$4$};
    \node at ($(v5)+(-0.25,-0.25)$) {$\bar{1}$};
    \node at ($(v6)+(-0.3,0)$) {$\bar{2}$};
    \node at ($(v7)+(-0.3,0)$) {$\bar{3}$};
    \node at ($(v8)+(-0.25,0.25)$) {$\bar{4}$};
    % diagonals with circle nodes at midpoints
    \draw[blue] (v3) -- (v1);
    \draw[OliveGreen] (v4) -- (v1);
    \draw[red] (v5) -- (v1);
    \draw[OliveGreen] (v5) -- (v8);
    \draw[blue] (v5) -- (v7);
    % arrows like the picture
    \end{tikzpicture}.
\end{align}
This is the centrally symmetric triangulation built from \textcolor{red}{$\Delta_{\bar{1}1}$}, \textcolor{blue}{$\Delta_{13}$} and \textcolor{OliveGreen}{$\Delta_{14}$} together with their images under rotation by $180^\circ$.

The chosen colours already illustrate the map between these two geometric descriptions:
\begin{align}
    \tilde\Delta_{ij} &\leftrightarrow (\Delta_{ij}=\Delta_{\bar{i}\bar{j}}) && \text{for }i<j\,,\nonumber\\
    \tilde\Delta_{ij}= 1-\tilde\Delta_{ji} &\leftrightarrow (\Delta_{\bar{j}i}=\Delta_{j\bar{i}}) && \text{for }i>j\,,\\
    \tilde\Delta_{iP} &\leftrightarrow (\Delta_{\bar{i}i})\,, &&\nonumber
\end{align}
where the brackets indicate a chord and its centrally symmetric image in the second description. Under this map, we see that all coordinates and compatibility relations in the punctured $(m+1)$-gon picture are equivalent to those in the $(2m+2)$-gon picture. Throughout most of the paper we use the $(2m+2)$-gon picture, but the punctured $(m+1)$-gon picture makes a brief appearance in Section~\ref{sec:sixpointtwo}.

%%%%%%%%%%%%%%%%%%%%%%%%%%%%%%%%%%%%%%%%%%%%%%%%%%%%%%%%%
%%%%%%%%%%%%%%%%%%%%%%%%%%%%%%%%%%%%%%%%%%%%%%%%%%%%%%%%%

\section{The General Coproduct Formula for Multiple Polylogarithms}

Here we review the coproduct of multiple polylogarithm functions. We use the definition of the depth-$k$ function $\li_{m_1,m_2,\ldots,m_k}(z_1,z_2,\ldots,z_k)$ given in~\eqref{eq:Li-convention}. For each index $m_i=1$, define the following lower-depth functions $\Li_i^{\mathrm L},\Li_i^{\mathrm R}$:
\begin{align}
 \Li_i^{\mathrm L}:={}&
 \begin{cases}
  \Li_{m_2,\ldots,m_k}(z_2,\ldots,z_k)\,,&i=1\,,\\
  \Li_{m_1,\ldots,m_{i-1},m_{i+1},\ldots,m_k}
  (z_1,\ldots,z_{i-2},z_{i-1}z_i,z_{i+1},\ldots,z_k)\,,&i>1\,,
 \end{cases}
 \label{eq:Li-left-merge}\\
 \Li_i^{\mathrm R}:={}&
 \Li_{m_1,\ldots,m_{i-1},m_{i+1},\ldots,m_k}
 (z_1,\ldots,z_{i-1},z_i z_{i+1},z_{i+2},\ldots,z_k)\,,
 \qquad i<k\,.
 \label{eq:Li-right-merge}
\end{align}
When the index list is empty, $\Li_{\varnothing}=1$.  With $N = m_1 + \cdots + m_k$, the general
$(N-1,1)$ coproduct component is~\cite{Duhr:2012fh}
\begin{equation}
\begin{aligned}
\delta_{N-1,1}
\Li_{m_1,\ldots,m_k}(z_1,\ldots,z_k)
={}&-\sum_{\substack{1\le i\le k\\1<m_i}}
\Li_{m_1,\ldots,m_i-1,\ldots,m_k}(z_1,\ldots,z_k)
\otimes\Li_1(1-z_i)\\
&+\sum_{\substack{1\le i\le k\\m_i=1}}
\Li_i^{\mathrm L}\otimes\Li_1(z_i)\\
&+\sum_{\substack{1\le i<k\\m_i=1}}
\Li_i^{\mathrm R}\otimes
\bigl[\Li_1(1-z_i)-\Li_1(z_i)\bigr]\,.
\end{aligned}
\label{eq:general-Li-coproduct}
\end{equation}

\section{The Coproduct of \texorpdfstring{$\qli^B$}{QLiB}}\label{sec:Bcancel}
In this section, we prove that the $\qli^B$ function defined in this paper satisfies the same differential equation~(\ref{eq:QLiB_differential_eq}) as the one~(\ref{eq:dgon}) satisfied by  $\psi_n^\circ$.

First, in Appendix \ref{sec:QLi_differential_match}, we show that the $i=1$ term in the second sum of \eqref{eq:general-Li-coproduct} exactly reproduces the differential equation for $\qli^B$ in \eqref{eq:QLiB_differential_eq}. We define the shorthand notation $E$ to encode the action of the $i=1$ term in the second sum:
\begin{equation}\label{eq:E_definition}
    E \Li_{m_1,\ldots,m_k}(z_1,\ldots,z_k) := \mathbf{1}_{1,m_1}\li_1^{\text{L}}\otimes\li_1(z_1)\,,
\end{equation}
where $\mathbf{1}_{a,b}=1$ when $a=b$ and 0 otherwise.
Second, in Appendices \ref{sec:Two_lemmas}, \ref{sec:Annihilation_branches} and \ref{sec:Annihilation_central_polygon}, we show that the remaining contribution in \eqref{eq:general-Li-coproduct} annihilates $\qli^B$
\begin{equation}
\label{eq:word-target}
    K \qli_n^B(1,\dots,2n,\bar{1},\dots,\overline{2n}) = 0\,.
\end{equation}
Here we use the shorthand notation for the difference $K=\delta_{N-1,1}-E$ whose action on $\li$ is given by
\begin{equation}
\begin{aligned}
    \label{eq:K_definition}
    K \Li_{m_1,\ldots,m_k}(z_1,\ldots,z_k)
    :=&-\sum_{\substack{1\le i\le k\\1<m_i}}
    \Li_{m_1,\ldots,m_i-1,\ldots,m_k}(z_1,\ldots,z_k)
    \otimes\Li_1(1-z_i)\\
    &+\sum_{\substack{1\color{red}<\color{black} i\le k\\m_i=1}}
    \Li_i^{\mathrm L}\otimes\Li_1(z_i)\\
    &+\sum_{\substack{1\le i<k\\m_i=1}}
    \Li_i^{\mathrm R}\otimes
    \bigl[\Li_1(1-z_i)-\Li_1(z_i)\bigr]\,.
\end{aligned}
\end{equation}
We highlight the small difference between $\delta_{N-1,1}$ and $K$ in red on the second line.
For notational convenience, we also define $K$ and $E$ to act directly on the weighted-word $W=[z_1,m_1]\otimes\cdots\otimes[z_k,m_k]$ via
\begin{equation}
\begin{aligned}
    &K W
    ={-}\sum_{\substack{1\le i\le k\\1<m_i}}W^0_{i}\otimes\li_1(1{-}z_i) {+}\sum_{\substack{1< i\le k\\m_i=1}}W^{\mathrm L}_{i}\otimes\li_1(z_i) {+}\sum_{\substack{1\le i<k\\m_i=1}}
    W^{\mathrm R}_{i}\otimes (\li_1(1{-}z_i){-}\li_1(z_i))\,,\\
    &E W=\mathbf{1}_{1,m_1}W_1^{\mathrm L}\otimes \li_1(z_1)\,,
    \label{eq:general-word-coproduct}
\end{aligned}
\end{equation}
where
\begin{equation}
\begin{aligned}
    &W^0_{i} = [z_1,m_1]\otimes\cdots[z_i,m_i-1]\cdots\otimes[z_k,m_k]\,,\\
    &W^{\mathrm L}_{i} = 
    \begin{cases}
        [z_2,m_2]\otimes\cdots\otimes[z_k,m_k]\quad  &i=1\,,\\
        [z_1,m_1]\otimes\cdots [z_{i-1}z_i,m_{i-1}] \cdots\otimes[z_k,m_k]\quad  &i>1\,,
    \end{cases}\\
    &W^{\mathrm R}_{i} = [z_1,m_1]\otimes\cdots [z_{i}z_{i+1},m_{i+1}] \cdots\otimes[z_k,m_k]\,.
\end{aligned}
\end{equation}
This is equivalent to \eqref{eq:K_definition} and \eqref{eq:E_definition} under the $\li$ map~(\ref{eq:Li0R}) and we use this notation to complete the proof.

Finally, we find it useful to combine \eqref{eq:tQB} and \eqref{eq:Tarpp} and re-express the weighted symbol of the $B$-type arborification in the form
\begin{equation}\label{eq:BFC}
    T_{\mathbf P}^B=(-1)^n\sum_{\mathbf{S}\subseteq \mathbf{P}_{4n}} F_\mathbf{S}\otimes\left[\crB(\mathbf{S}),|\mathbf{S}|/4\right],
\end{equation}
where $\mathbf{S}$ is a central polygon and $F_{\mathbf{S}}$ is the quasi-shuffle product of all branches attached to the root in the dual tree. 

\subsection{Matching the Differential Equation}
\label{sec:QLi_differential_match}
In this section, we prove that $E$ acting on $\qli_n^B$ reproduces the differential equation \eqref{eq:QLiB_differential_eq}. According to \eqref{eq:general-word-coproduct}, $E$ only picks out the weight-one first entry of a weighted symbol; therefore, we obtain the form
\begin{equation}
    E (T_{\mathbf{P}}^B) = \sum_{i=1}^{2n} W_i \otimes \li_1((q_{\mathbf{P}(i)\,\mathbf{P}(i+1)\,\mathbf{P}(i+2)\,\mathbf{P}(i+3)})^{(-1)^{\pi(\mathbf{P}(i))}})\,,
\end{equation}
where $W_i$ is a weighted symbol of weight-$(n{-}1)$. Let us focus on the contribution to a particular term in the summand, namely $W_i\otimes \li_1((q_{\mathbf{P}(i)\,\mathbf{P}(i+1)\,\mathbf{P}(i+2)\,\mathbf{P}(i+3)})^{(-1)^{\pi(\mathbf{P}(i))}})$. The contributing quadrangulations are all partial quadrangulations that explicitly cut out $(\mathbf{P}(i),\mathbf{P}(i+1),\mathbf{P}(i+2),\mathbf{P}(i+3))$. In other words, the contributing quadrangulations are the full set of partial quadrangulations of 
\begin{multline}
    \mathbf{P}_{\widehat{i{+}1}, \widehat{i{+}2}} =\\
    (\mathbf{P}(1),\cdots,\widehat{\mathbf{P}(i{+}1)},\widehat{\mathbf{P}(i{+}2)},\cdots,\mathbf{P}(2n),\overline{\mathbf{P}(1)},\cdots,\widehat{\overline{\mathbf{P}(i{+}1)}},\widehat{\overline{\mathbf{P}(i{+}2)}},\cdots,\overline{\mathbf{P}(2n)})\,,
\end{multline}
appending a quadrilateral $(\mathbf{P}(i),\mathbf{P}(i+1),\mathbf{P}(i+2),\mathbf{P}(i+3))$. In the construction of the weighted symbol, the contributing terms select, from the quasi-shuffle product, the tensor-product term that places the weighted letter $[(q_{\mathbf{P}(i)\,\mathbf{P}(i+1)\,\mathbf{P}(i+2)\,\mathbf{P}(i+3)})^{(-1)^{\pi(\mathbf{P}(i))}},1]$ in the first entry. Summing up all contributing quadrangulations, we obtain exactly
\begin{equation}
    (-1)^{\pi(\mathbf{P}(i))+1} T^B_{\mathbf{P}_{\widehat{i{+}1},\widehat{i{+}2}}} \otimes \li_1((q_{\mathbf{P}(i)\,\mathbf{P}(i+1)\,\mathbf{P}(i+2)\,\mathbf{P}(i+3)})^{(-1)^{\pi(\mathbf{P}(i))}})\,.
\end{equation}
Using the definition of $\qli_1^\pm$ in \eqref{eq:qliA1} and summing over all $i$, we obtain
\begin{equation}
    E(T_{\mathbf{P}}^B) = \sum_{i=1}^{2n} T^B_{\mathbf{P}_{\widehat{i{+}1},\widehat{i{+}2}}}\otimes \qli_1^{(-1)^{\pi(\mathbf{P}(i))}}(\mathbf{P}(i),\mathbf{P}(i+1),\mathbf{P}(i+2),\mathbf{P}(i+3))\,.
\end{equation}
Under the $\li$ map~(\ref{eq:Li0R}), this is exactly \eqref{eq:QLiB_differential_eq}.

\subsection{Two Lemmas for Weighted Words}\label{sec:Two_lemmas}

In this section, we prove two lemmas that are instrumental in proving $K \qli_n^B =0$. The first lemma is a reorganisation of the weighted symbol of the branches, with the dot product in the even quadrilateral explicitly expanded in \eqref{eq:Aarbe}. The second lemma states that the operator $K$ acting on weighted symbols satisfies the Leibniz rule.

\subsubsection{Rewriting branches}
\begin{figure}
\centering
\color{orange}
\begin{tikzpicture}[scale=0.49, line cap=round, line join=round]

% radii
\def\R{3}
\def\Rout{3.4}   % radius for dots outside the circle
\def\rDot{0.045}  % size of dots

% circle
\draw[thick] (0,0) circle (\R);

% four symmetric points on the circle
\coordinate (LT) at (110:\R);  % left top
\coordinate (RT) at ( 70:\R);  % right top
\coordinate (LB) at (230:\R);  % left bottom
\coordinate (RB) at (310:\R);  % right bottom

% left chord: bends inward
\draw[thick]
(LT) .. controls (-0.85, 1.0) and (-0.85,-1.0) .. (LB);

% right chord: bends inward
\draw[thick]
(RT) .. controls ( 0.85, 1.0) and ( 0.85,-1.0) .. (RB);

% dots outside the boundary
% right arc
\foreach \ang in {10,0,-10}
  \fill (\ang:\Rout) circle (\rDot);

% left arc
\foreach \ang in {190,180,170}
  \fill (\ang:\Rout) circle (\rDot);

% labels
\node[above left=2pt]  at (LT) {$\mathbf{Q}(i)$};
\node[below left=2pt]  at (LB) {$\mathbf{Q}(1)$};
\node[above right=2pt] at (RT) {$\mathbf{Q}(i{+}1)$};
\node[below right=2pt] at (RB) {$\mathbf{Q}(|\mathbf{Q}|)$};
\end{tikzpicture}
\color{brown}
\begin{tikzpicture}[scale=0.49, line cap=round, line join=round]

% radii
\def\R{3}
\def\Rout{3.4}   % radius for dots outside the circle
\def\rDot{0.045}  % size of dots

% circle
\draw[thick] (0,0) circle (\R);

% four symmetric points on the circle
\coordinate (LT) at (130:\R);  % left top
\coordinate (RT) at ( 50:\R);  % right top
\coordinate (LB) at (230:\R);  % left bottom
\coordinate (RB) at (310:\R);  % right bottom

% side chords
\draw[thick]
(LT) .. controls (-1.55, 1.0) and (-1.55,-1.0) .. (LB);

\draw[thick]
(RT) .. controls ( 1.55, 1.0) and ( 1.55,-1.0) .. (RB);

% top chord
\draw[thick]
(LT) .. controls (-0.9,1.8) and (0.9,1.8) .. (RT);

% dots outside the boundary
% top arc
\foreach \ang in {100,90,80}
  \fill (\ang:\Rout) circle (\rDot);

% right arc
\foreach \ang in {10,0,-10}
  \fill (\ang:\Rout) circle (\rDot);

% left arc
\foreach \ang in {190,180,170}
  \fill (\ang:\Rout) circle (\rDot);

% labels
\node[above left=2pt]  at (LT) {$\mathbf{Q}(i)$};
\node[below left=2pt]  at (LB) {$\mathbf{Q}(1)$};
\node[above right=2pt] at (RT) {$\mathbf{Q}(i{+}3)$};
\node[below right=2pt] at (RB) {$\mathbf{Q}(|\mathbf{Q}|)$};
\end{tikzpicture}
\color{black}
\caption{Two classes contributing to the rooted polygon $\mathbf{Q}$; these have support only on even $i$.}
\label{fig:rewrite_branch}
\end{figure}
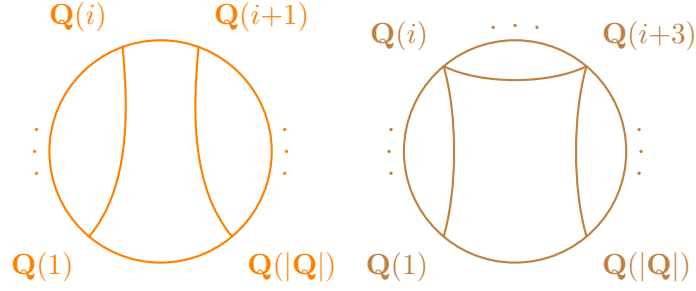

For an alternating polygon $\mathbf{R}=(\mathbf{R}(1),\dots,\mathbf{R}(|\mathbf{R}|))$ with the root edge $(\mathbf{R}(1),\mathbf{R}(|\mathbf{R}|))$ that does not contain a central polygon, we claim the weighted symbol derived from arborification can be rewritten as 
\begin{equation}
\begin{aligned}
    \label{eq:complete-last-letter}
    &T^A_\mathbf{R}=
    \begin{cases}
    \displaystyle\sum_{\substack{\mathbf{Q}\subseteq \mathbf{R}\\ |\mathbf{Q}|\ge4}}
    \left(\star_jT_{j}^A\right)
    \otimes[q_{\mathbf{Q}},|\mathbf{Q}|/2-1]\,,&\pi(\mathbf{R}(1))\text{ even}\,,\\[4mm]
    \displaystyle-\sum_{\substack{\mathbf{Q}\subseteq \mathbf{R}\\ |\mathbf{Q}|=4}}
    \left(\star_jT_{j}^A\right)
    \otimes[1/q_{\mathbf{Q}},1]\,,&\pi(\mathbf{R}(1))\text{ odd}\,,
    \end{cases}\\
    &q_{\mathbf{Q}}=\prod_{1\leq i\leq \frac{|\mathbf{Q}|-2}{2}} q_{\mathbf{Q}(1)\,\mathbf{Q}(2i)\,\mathbf{Q}(2i+1)\,\mathbf{Q}(2i+2)}
\end{aligned}
\end{equation}
where $\mathbf{Q}$ runs over the root polygons that contain the root edge ($\mathbf{R}(1)$, $\mathbf{R}(|\mathbf{R}|)$). For $\pi(\mathbf{R}(1))$ odd, this is obtained by summing \eqref{eq:Aarbo} over all possible root quadrilaterals. For $\pi(\mathbf{R}(1))$ even, we prove it by induction. The base case, $|\mathbf{R}|=4$, is obviously true. Assume \eqref{eq:complete-last-letter} is true for $|\mathbf{R}|\leq 2 R^*$. Then for $|\mathbf{R}|=2R^*{+}2$, the weighted symbol is 
\begin{equation}
\begin{aligned}
    T^A_\mathbf{R}=&\sum_{1\leq i\leq j\leq R^*} \left(T^A_{\mathbf{R}(1),\cdots,\mathbf{R}(2i)}\star T^A_{\mathbf{R}(2i),\cdots,\mathbf{R}(2j{+}1)}\star T^A_{\mathbf{R}(2j{+}1),\cdots,\mathbf{R}(2R^*+2)}\right)\\
    &\hspace{8cm}\otimes [q_{\mathbf{R}(1)\,\mathbf{R}(2i)\,\mathbf{R}(2j{+}1)\,\mathbf{R}(2R^*+2)},1]\\
    &+\sum_{1\leq i\leq j\leq R^*} \left(T^A_{\mathbf{R}(1),\cdots,\mathbf{R}(2i)}\star T^A_{\mathbf{R}(2i),\cdots,\mathbf{R}(2j{+}1)}\star T^A_{\mathbf{R}(2j+1),\cdots,\mathbf{R}(2R^*+2)}\right)\\
    &\hspace{8cm}\cdot [q_{\mathbf{R}(1)\,\mathbf{R}(2i)\,\mathbf{R}(2j{+}1)\,\mathbf{R}(2R^*+2)},1]\,,
\end{aligned}
\end{equation}
where each of the sub-branches is given by \eqref{eq:complete-last-letter}. The first line is exactly the $|\mathbf{Q}|=4$ case of the summand in \eqref{eq:complete-last-letter} and the second line generates all cases with $|\mathbf{Q}|\geq 6$. 
The contributions to a particular alternating polygon with $|\mathbf{Q}|\geq 6$ are
\begin{equation}
    \left(\star_jT^A_{j}\right)
    \otimes[q_{\mathbf{Q}},|\mathbf{Q}|/2-1]\left[\color{orange}\left(\frac{|\mathbf{Q}|}{2}-1\right)\color{brown}-\left(\frac{|\mathbf{Q}|}{2}-2\right)\color{black}\right] = \left(\star_jT^A_{j}\right)
    \otimes[q_{\mathbf{Q}},|\mathbf{Q}|/2-1]\,.
\end{equation}
The two integers count terms in the two contributing classes, illustrated in Figure \ref{fig:rewrite_branch}. The boundary polygons in Figure \ref{fig:rewrite_branch} are the root polygons of the corresponding sub-branches. Since $\mathbf{R}(2i)$ is odd, the root polygon of the middle sub-branch can either contribute a quadrilateral or nothing. The class with a quadrilateral has one fewer term than the class with no contribution from the middle branch.

\subsubsection{The Leibniz rule and the letter appending formula}
Under the $\Li$ map~(\ref{eq:Li0R}), the quasi-shuffle product becomes ordinary multiplication, $\Li(U\star V)=\Li(U)\Li(V)$~\cite{hoffman1999quasishuffleproducts}. Therefore, the coproduct $\delta_{N-1,1}$, which is equivalent to the differential, obeys the Leibniz rule when acting on the quasi-shuffle product. The same is true for $E$: in \eqref{eq:BstardefB}, one tensor product term begins with the first letter of $U$ and the other with the first letter of $V$, whereas the dot product begins with a letter of weight greater than one and has no support under $E$. Thus we obtain
\begin{equation}
 K(U\star V)
 =\bigl(K U\bigr)\star V
  +U\star\bigl(K V\bigr)\,.
 \label{eq:KLeibniz}
\end{equation}

Now consider a nonempty weighted symbol term $W=V\otimes[q_{S},m]$ and append an additional weighted letter on the right to obtain $W\otimes [q_d,m']$. The action of $K$ on $W\otimes [q_d,m']$ produces three new terms in addition to $K(W)\otimes [q_d,m']$. We call them the \color{blue}weight-lowering\color{black}, \color{brown}left deletion \color{black} and \color{red}right deletion \color{black} terms, respectively:
\begin{equation}
\begin{aligned}
    K\bigl(W\otimes[q_d,m']\bigr)
    ={}(K(W))\otimes[q_d,m']
    &\color{blue}+\theta(m'-1)\bigl(W\otimes[q_d,m'-1]\bigr)\otimes\log q_d\\
    &\color{brown}-\delta_{m',1}\bigl(V\otimes[q_d q_{S},m]\bigr)\otimes\log(1-q_d)\\
    &\color{red}+\delta_{m,1}\bigl(V\otimes[q_d q_{S},m']\bigr)
       \otimes\log\frac{1-q_{S}}{q_{S}}\,\
\label{eq:append}
\end{aligned}
\end{equation}
where $\theta(m'-1)=1$ if $m^\prime>1$ and vanishes if $m^\prime=1$.

The appending formula~(\ref{eq:append}) for the coproduct is used repeatedly in our proof, including in our analysis of the action of $K$ on the weighted symbol given in \eqref{eq:BFC} and \eqref{eq:complete-last-letter}. We prove \eqref{eq:word-target} in two steps. First, we prove that $K$ annihilates the quasi-shuffle product of all branches in the arborification algorithm
\begin{equation}
    K(F_{\mathbf{R}}) =0\,.
\end{equation}
Next, we prove the total contribution from left deletion, right deletion and weight-lowering terms generated from appending the central polygon symbol $[\crB(\mathbf{S}),|\mathbf{S}|/4]$ to $F_{\mathbf{S}}$ vanishes.

\subsection{Annihilation of Branches}\label{sec:Annihilation_branches}
In this section, we prove that the quasi-shuffle product of all branches in \eqref{eq:BFC} is annihilated by $K$. Since the action of $K$ on the quasi-shuffle product satisfies the Leibniz rule, it suffices to show that
\begin{equation}
    K (T^A_{\mathbf{R}}) =0
\end{equation}
for every branch $\mathbf{R}$ in the dual tree (equivalently, every alternating polygon that does not contain the central polygon), where $T^A_{\mathbf{R}}$ denotes its weighted symbol. We prove the claim by induction. The base case, a weight-one quadrilateral, is obviously true. Suppose that $K (T^A_{\mathbf{R}'}) =0$ for every weighted symbol $T^A_{\mathbf{R}'}$ of weight at most $R^*$. Now consider a weighted symbol $T^A_{\mathbf{R}}$ of weight $R^*+1$. According to \eqref{eq:complete-last-letter} and \eqref{eq:append}, applying $K$ to $T^A_{\mathbf{R}}$ yields the form
\begin{equation}
%\begin{aligned}
\label{eq:K_act_on_TR}
    K(T^A_\mathbf{R}) = \sum_{\substack{\mathbf{Q}\subseteq \mathbf{R}\\ 6\leq|\mathbf{Q}|\leq|\mathbf{R}|}} (\star_j T^A_j)\otimes [(q_{\mathbf{Q}})^{(-1)^{\pi(\mathbf{Q}(1)})},|\mathbf{Q}|/2-2]\otimes [\text{final entry}]\,,\\
    % &q_{S}=\prod_{1\leq i\leq \frac{|S|}{2}-1} (q_{S(1)\,S(2i)\,S(2i+1)\,S(2i+2)})^{(-1)^{S(1)}},
%\end{aligned}
\end{equation}
where $\mathbf{Q}$ is an alternating polygon containing the root edge $(\mathbf R(1),\mathbf R(|\mathbf R|)){=}(\mathbf Q(1),\mathbf Q(|\mathbf Q|))$ and $q_{\mathbf{Q}}$ is defined in \eqref{eq:complete-last-letter}. The analysis splits into two cases according to the parity of $R$. 

\subsubsection{Parity odd}
\begin{figure}
\centering
\color{orange}
\begin{tikzpicture}[scale=0.49, line cap=round, line join=round]

% radii
\def\R{3}
\def\Rout{3.4}   % radius for dots outside the circle
\def\rDot{0.045}  % size of dots

% circle
\draw[thick] (0,0) circle (\R);

% four symmetric points on the circle
\coordinate (LT) at (130:\R);  % left top
\coordinate (RT) at ( 50:\R);  % right top
\coordinate (LB) at (230:\R);  % left bottom
\coordinate (RB) at (310:\R);  % right bottom

% side chords
\draw[thick]
(LT) .. controls (-1.55, 1.0) and (-1.55,-1.0) .. (LB);

\draw[thick]
(RT) .. controls ( 1.55, 1.0) and ( 1.55,-1.0) .. (RB);

% top chord
\draw[thick]
(LT) .. controls (-0.9,1.8) and (0.9,1.8) .. (RT);

% dots outside the boundary
% top arc
\foreach \ang in {100,90,80}
  \fill (\ang:\Rout) circle (\rDot);

% right arc
\foreach \ang in {10,0,-10}
  \fill (\ang:\Rout) circle (\rDot);

% left arc
\foreach \ang in {190,180,170}
  \fill (\ang:\Rout) circle (\rDot);

% labels
\node[above left=2pt]  at (LT) {$\mathbf Q(4)$};
\node[below left=2pt]  at (LB) {$\mathbf Q(1)$};
\node[above right=2pt] at (RT) {$\mathbf Q(|\mathbf Q|{-}3)$};
\node[below right=2pt] at (RB) {$\mathbf Q(|\mathbf Q|)$};

\end{tikzpicture}
\color{green}
\begin{tikzpicture}[scale=0.49, line cap=round, line join=round]

% radii
\def\R{3}
\def\Rout{3.4}
\def\rDot{0.045}

% circle
\draw[thick] (0,0) circle (\R);

% boundary points
\coordinate (LT) at (130:\R);   % S(4)
\coordinate (LB) at (230:\R);   % S(1)
\coordinate (RM) at (-30:\R);    % S(2|S|-1)
\coordinate (RB) at (270:\R);   % S(2|S|)

% left chord
\draw[thick]
(LT) .. controls (-1.55,1.0) and (-1.55,-1.0) .. (LB);

% top chord: bows inward
\draw[thick]
(LT) .. controls (-0.8,0.0) and (0.6,-1.1) .. (RM);

% dots outside the boundary
% top arc
\foreach \ang in {75,60,45,30,15}
  \fill (\ang:\Rout) circle (\rDot);

% left arc
\foreach \ang in {170,180,190}
  \fill (\ang:\Rout) circle (\rDot);

% labels
\node[above left=2pt] at (LT) {$\mathbf Q(4)$};
\node[below left=2pt] at (LB) {$\mathbf Q(1)$};
\node[below right=4pt] at (RM) {$\mathbf Q(|\mathbf Q|{-}1)$};
\node[below =2pt] at (RB) {$\mathbf Q(|\mathbf Q|)$};

\end{tikzpicture}
\color{brown}
\begin{tikzpicture}[scale=0.49, line cap=round, line join=round]

% radii
\def\R{3}
\def\Rout{3.4}   % radius for dots outside the circle
\def\rDot{0.045}  % size of dots

% circle
\draw[thick] (0,0) circle (\R);

% four symmetric points on the circle
\coordinate (LT) at (210:\R);  % left top
\coordinate (RT) at ( 50:\R);  % right top
\coordinate (LB) at (270:\R);  % left bottom
\coordinate (RB) at (310:\R);  % right bottom

% side chords

\draw[thick]
(RT) .. controls ( 1.55, 1.0) and ( 1.55,-1.0) .. (RB);

% top chord
\draw[thick]
(LT) .. controls (0.0,-0.9) and (1.15,-0.05) .. (RT);

% dots outside the boundary
% top arc
\foreach \ang in {105,120,135,150,165}
  \fill (\ang:\Rout) circle (\rDot);

% right arc
\foreach \ang in {10,0,-10}
  \fill (\ang:\Rout) circle (\rDot);

% labels
\node[below left=2pt]  at (LT) {$\mathbf Q(2)$};
\node[below =2pt]  at (LB) {$\mathbf Q(1)$};
\node[above right=2pt] at (RT) {$\mathbf Q(|\mathbf Q|{-}3)$};
\node[below right=2pt] at (RB) {$\mathbf Q(|\mathbf Q|)$};

\end{tikzpicture}
\color{cyan}
\begin{tikzpicture}[scale=0.49, line cap=round, line join=round]

% radii
\def\R{3}
\def\Rout{3.4}   % radius for dots outside the circle
\def\rDot{0.045}  % size of dots

% circle
\draw[thick] (0,0) circle (\R);

% four symmetric points on the circle
\coordinate (LT) at (200:\R);  % left top
\coordinate (RT) at ( -20:\R);  % right top
\coordinate (LB) at (230:\R);  % left bottom
\coordinate (RB) at (310:\R);  % right bottom

% top chord
\draw[thick]
(LT) .. controls (-1.1,-0.65) and (1.1,-0.65) .. (RT);

% dots outside the boundary
% top arc
\foreach \ang in {50,70,90,110,130}
  \fill (\ang:\Rout) circle (\rDot);

% labels
\node[below left=2pt]  at (LT) {$\mathbf Q(2)$};
\node[below =6pt]  at (LB) {$\mathbf Q(1)$};
\node[below right =2pt] at (RT) {$\mathbf Q(|\mathbf Q|{-}1)$};
\node[below =7pt] at (RB) {$\mathbf Q(|\mathbf Q|)$};

\end{tikzpicture}
\color{black}
\caption{The four contributions to the cancellation of polygon $\mathbf Q$.}
\label{fig:Odd_r_octagon_and_above}
\end{figure}
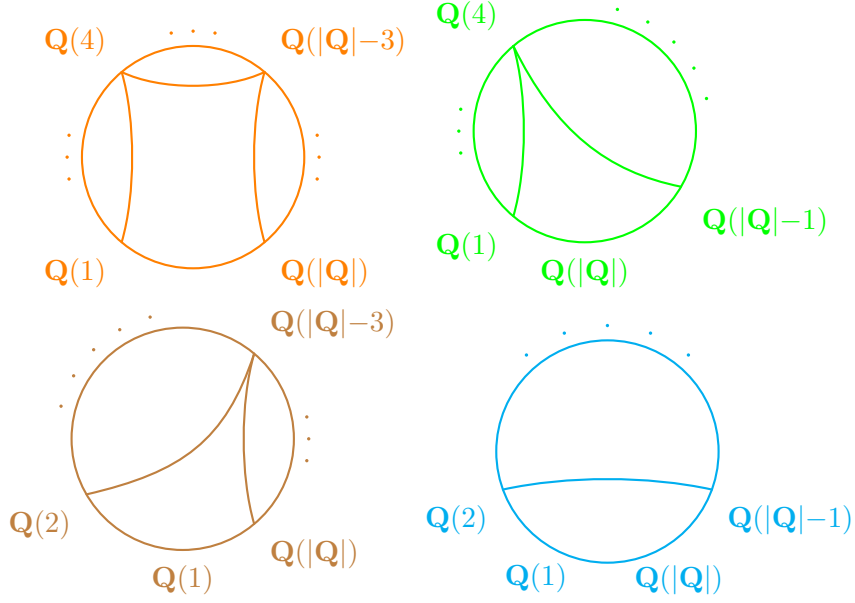

The summand in \eqref{eq:K_act_on_TR} only receives contributions from left and right deletion in \eqref{eq:append} but not from the weight-lowering term. The reason is simply that the last entry of parity-odd $T^A_{\mathbf R}$ is always of weight one. We sum up all contributions to a particular alternating polygon $\mathbf Q$ and show that the final entry cancels. For $|\mathbf Q|\geq 8$, the contributing terms for each polygon $\mathbf Q$ cancel:
\begin{multline}
    (\star_j T^A_j)\otimes [1/q_{\mathbf Q},|\mathbf Q|/2-2] \otimes \bigg[\color{orange}\log(q_{\mathbf Q(4)\,\mathbf Q(|\mathbf Q|)\,\mathbf Q(|\mathbf Q|-3)\,\mathbf Q(1)})\color{green}-\log(q_{\mathbf Q(4)\,\mathbf Q(|\mathbf Q|)\,\mathbf Q(|\mathbf Q|-1)\,\mathbf Q(1)})\\
    \color{brown}-\log(q_{\mathbf Q(2)\,\mathbf Q(|\mathbf Q|)\,\mathbf Q(|\mathbf Q|-3)\,\mathbf Q(1)})\color{cyan}+\log(q_{\mathbf Q(2)\,\mathbf Q(|\mathbf Q|)\,\mathbf Q(|\mathbf Q|-1)\,\mathbf Q(1)})\color{black}\bigg] =0\,.
\end{multline}
The four terms arise from left deletion in \eqref{eq:append} for the four contributions illustrated in Figure \ref{fig:Odd_r_octagon_and_above}, while there are no contributions from right deletion.

\begin{figure}
\centering
\color{red}
\begin{tikzpicture}[scale=0.5, line cap=round, line join=round]
\def\R{2.5}

% vertices
\coordinate (1) at (240:\R);
\coordinate (2) at (180:\R);
\coordinate (3) at (120:\R);
\coordinate (4) at ( 60:\R);
\coordinate (5) at (  0:\R);
\coordinate (6) at (300:\R);

% hexagon
\draw[thick] (1)--(2)--(3)--(4)--(5)--(6)--cycle;

% straight line from corner 2 to corner 5
\draw[thick] (2)--(5);

% labels
\node[below left]  at (1) {$\mathbf Q(1)$};
\node[left]        at (2) {$\mathbf Q(2)$};
\node[above left]  at (3) {$\mathbf Q(3)$};
\node[above right] at (4) {$\mathbf Q(4)$};
\node[right]       at (5) {$\mathbf Q(5)$};
\node[below right] at (6) {$\mathbf Q(6)$};
\end{tikzpicture}
\color{brown}
\begin{tikzpicture}[scale=0.5, line cap=round, line join=round]
\def\R{2.5}

% vertices
\coordinate (1) at (240:\R);
\coordinate (2) at (180:\R);
\coordinate (3) at (120:\R);
\coordinate (4) at ( 60:\R);
\coordinate (5) at (  0:\R);
\coordinate (6) at (300:\R);

% hexagon
\draw[thick] (1)--(2)--(3)--(4)--(5)--(6)--cycle;

% straight line from corner 2 to corner 5
\draw[thick] (3)--(6);

% labels
\node[below left]  at (1) {$\mathbf Q(1)$};
\node[left]        at (2) {$\mathbf Q(2)$};
\node[above left]  at (3) {$\mathbf Q(3)$};
\node[above right] at (4) {$\mathbf Q(4)$};
\node[right]       at (5) {$\mathbf Q(5)$};
\node[below right] at (6) {$\mathbf Q(6)$};
\end{tikzpicture}
\color{blue}
\begin{tikzpicture}[scale=0.5, line cap=round, line join=round]
\def\R{2.5}

% vertices
\coordinate (1) at (240:\R);
\coordinate (2) at (180:\R);
\coordinate (3) at (120:\R);
\coordinate (4) at ( 60:\R);
\coordinate (5) at (  0:\R);
\coordinate (6) at (300:\R);

% hexagon
\draw[thick] (1)--(2)--(3)--(4)--(5)--(6)--cycle;

% straight line from corner 2 to corner 5
\draw[thick] (1)--(4);

% labels
\node[below left]  at (1) {$\mathbf Q(1)$};
\node[left]        at (2) {$\mathbf Q(2)$};
\node[above left]  at (3) {$\mathbf Q(3)$};
\node[above right] at (4) {$\mathbf Q(4)$};
\node[right]       at (5) {$\mathbf Q(5)$};
\node[below right] at (6) {$\mathbf Q(6)$};
\end{tikzpicture}
\color{black}
\caption{The three contributions to the cancellation of hexagon $\mathbf Q$.}
\label{fig:Odd_r_hexagon}
\end{figure}
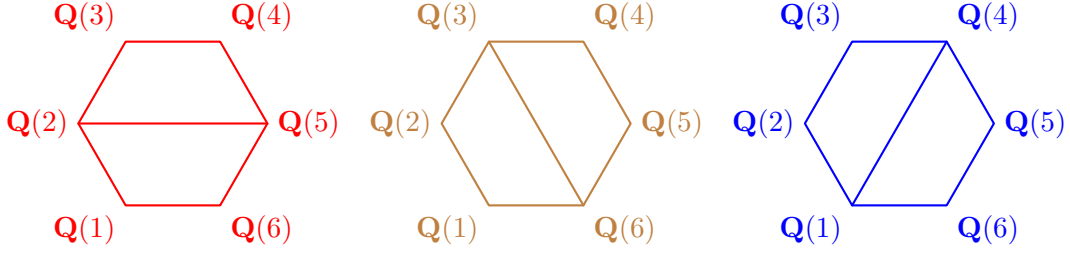

For the special case $|\mathbf Q|=6$, the contributing terms for a hexagon $\mathbf Q$ cancel,
\begin{equation}
\begin{aligned}
    (\star_j T^A_j)\otimes [1/q_\mathbf Q,1] \otimes &\bigg[\color{red}\log(q_{\mathbf Q(2)\,\mathbf Q(6)\,\mathbf Q(5)\,\mathbf Q(1)})+\log(q_{\mathbf Q(3)\,\mathbf Q(2)\,\mathbf Q(4)\,\mathbf Q(5)})\\
    &\color{brown}-\log(q_{\mathbf Q(2)\,\mathbf Q(6)\,\mathbf Q(3)\,\mathbf Q(1)})-\log(q_{\mathbf Q(4)\,\mathbf Q(5)\,\mathbf Q(3)\,\mathbf Q(6)})\\
    &\color{blue}-\log(q_{\mathbf Q(4)\,\mathbf Q(6)\,\mathbf Q(5)\,\mathbf Q(1)})-\log(q_{\mathbf Q(2)\,\mathbf Q(3)\,\mathbf Q(1)\,\mathbf Q(4)})\color{black}\bigg] =0\,.
\end{aligned}
\end{equation}
The six terms arise from the left and right deletion in \eqref{eq:append} for the three quadrangulations of hexagon $\mathbf Q$ illustrated in Figure \ref{fig:Odd_r_hexagon}.
This concludes the proof of cancellation of $K(T^A_{\mathbf R})$ for the parity-odd case.

\subsubsection{Parity even}
The summand in \eqref{eq:K_act_on_TR} receives contributions from left and right deletion, as well as lowering in \eqref{eq:append}, which is absent in the parity-odd case. For $|\mathbf Q|\geq 8$, the contributions for each polygon $\mathbf Q$ cancel.
\begin{equation}
\begin{aligned}
    &(\star_j T^A_j)\otimes [q_{\mathbf Q},|\mathbf Q|/2{-}2]\otimes \bigg[ \log(q_{\mathbf Q}) \\
    &\hspace{1mm}\color{orange}{-}\sum_{j=1}^{\frac{|\mathbf Q|}{2}{-}1}\log(q_{\mathbf Q(1)\,\mathbf Q(2j{+}1)\,\mathbf Q(2j)\,\mathbf Q(|\mathbf Q|)})\color{brown}+\sum_{j=1}^{\frac{|\mathbf Q|}{2}{-}2}\log(q_{\mathbf Q(1)\,\mathbf Q(2j{+}3)\,\mathbf Q(2j)\,\mathbf Q(|\mathbf Q|)})\\
    &\hspace{1mm}\color{red}-\sum_{j=1}^{\frac{|\mathbf Q|}{2}-1}\log(q_{\mathbf Q(2j)\,\mathbf Q(2j-1)\,\mathbf Q(2j+1)\,\mathbf Q(2j+2)})\color{blue}+\sum_{j=1}^{\frac{|\mathbf Q|}{2}-2}\log(q_{\mathbf Q(2j+2)\,\mathbf Q(2j+1)\,\mathbf Q(2j+3)\,\mathbf Q(2j)})\color{black}\bigg]=0\,.
\end{aligned}
\end{equation}
The two terms on the second line arise from left deletion, while the terms on the last line arise from right deletion. The corresponding quadrangulations are illustrated in Figure \ref{fig:Even_r_octagon_and_above}. The first piece arises from the weight-lowering contribution in \eqref{eq:append},
\begin{equation}
    \left.K((\star_j T^A_j)\otimes [q_{\mathbf Q},|\mathbf Q|/2-1])\right|_{\text{lowering}} = (\star_j T^A_j)\otimes [q_{\mathbf Q},|\mathbf Q|/2-2] \otimes \log(q_{\mathbf Q})\,.
\end{equation}
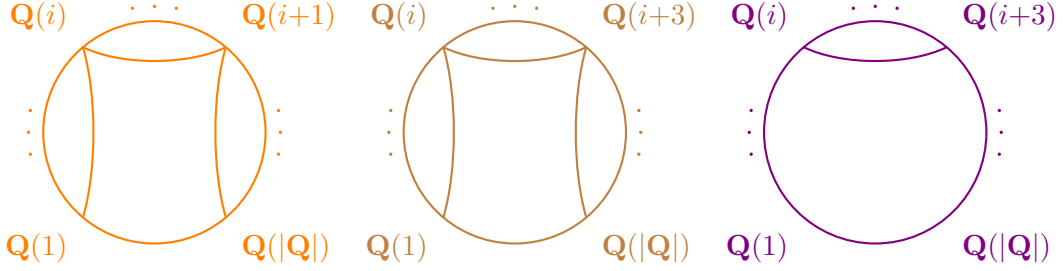
\begin{figure}
\centering
\color{orange}
\begin{tikzpicture}[scale=0.49, line cap=round, line join=round]

% radii
\def\R{3}
\def\Rout{3.4}   % radius for dots outside the circle
\def\rDot{0.045}  % size of dots

% circle
\draw[thick] (0,0) circle (\R);

% four symmetric points on the circle
\coordinate (LT) at (130:\R);  % left top
\coordinate (RT) at ( 50:\R);  % right top
\coordinate (LB) at (230:\R);  % left bottom
\coordinate (RB) at (310:\R);  % right bottom

% side chords
\draw[thick]
(LT) .. controls (-1.55, 1.0) and (-1.55,-1.0) .. (LB);

\draw[thick]
(RT) .. controls ( 1.55, 1.0) and ( 1.55,-1.0) .. (RB);

% top chord
\draw[thick]
(LT) .. controls (-0.9,1.8) and (0.9,1.8) .. (RT);

% dots outside the boundary
% top arc
\foreach \ang in {100,90,80}
  \fill (\ang:\Rout) circle (\rDot);

% right arc
\foreach \ang in {10,0,-10}
  \fill (\ang:\Rout) circle (\rDot);

% left arc
\foreach \ang in {190,180,170}
  \fill (\ang:\Rout) circle (\rDot);

% labels
\node[above left=2pt]  at (LT) {$\mathbf Q(i)$};
\node[below left=2pt]  at (LB) {$\mathbf Q(1)$};
\node[above right=2pt] at (RT) {$\mathbf Q(i{+}1)$};
\node[below right=2pt] at (RB) {$\mathbf Q(|\mathbf Q|)$};

\end{tikzpicture}
\color{brown}
\begin{tikzpicture}[scale=0.49, line cap=round, line join=round]

% radii
\def\R{3}
\def\Rout{3.4}   % radius for dots outside the circle
\def\rDot{0.045}  % size of dots

% circle
\draw[thick] (0,0) circle (\R);

% four symmetric points on the circle
\coordinate (LT) at (130:\R);  % left top
\coordinate (RT) at ( 50:\R);  % right top
\coordinate (LB) at (230:\R);  % left bottom
\coordinate (RB) at (310:\R);  % right bottom

% side chords
\draw[thick]
(LT) .. controls (-1.55, 1.0) and (-1.55,-1.0) .. (LB);

\draw[thick]
(RT) .. controls ( 1.55, 1.0) and ( 1.55,-1.0) .. (RB);

% top chord
\draw[thick]
(LT) .. controls (-0.9,1.8) and (0.9,1.8) .. (RT);

% dots outside the boundary
% top arc
\foreach \ang in {100,90,80}
  \fill (\ang:\Rout) circle (\rDot);

% right arc
\foreach \ang in {10,0,-10}
  \fill (\ang:\Rout) circle (\rDot);

% left arc
\foreach \ang in {190,180,170}
  \fill (\ang:\Rout) circle (\rDot);

% labels
\node[above left=2pt]  at (LT) {$\mathbf Q(i)$};
\node[below left=2pt]  at (LB) {$\mathbf Q(1)$};
\node[above right=2pt] at (RT) {$\mathbf Q(i{+}3)$};
\node[below right=2pt] at (RB) {$\mathbf Q(|\mathbf Q|)$};

\end{tikzpicture}
\color{violet}
\begin{tikzpicture}[scale=0.49, line cap=round, line join=round]

% radii
\def\R{3}
\def\Rout{3.4}   % radius for dots outside the circle
\def\rDot{0.045}  % size of dots

% circle
\draw[thick] (0,0) circle (\R);

% four symmetric points on the circle
\coordinate (LT) at (130:\R);  % left top
\coordinate (RT) at ( 50:\R);  % right top
\coordinate (LB) at (230:\R);  % left bottom
\coordinate (RB) at (310:\R);  % right bottom
% top chord
\draw[thick]
(LT) .. controls (-0.9,1.8) and (0.9,1.8) .. (RT);

% dots outside the boundary
% top arc
\foreach \ang in {100,90,80}
  \fill (\ang:\Rout) circle (\rDot);

% right arc
\foreach \ang in {10,0,-10}
  \fill (\ang:\Rout) circle (\rDot);

% left arc
\foreach \ang in {190,180,170}
  \fill (\ang:\Rout) circle (\rDot);

% labels
\node[above left=2pt]  at (LT) {$\mathbf Q(i)$};
\node[below left=2pt]  at (LB) {$\mathbf Q(1)$};
\node[above right=2pt] at (RT) {$\mathbf Q(i{+}3)$};
\node[below right=2pt] at (RB) {$\mathbf Q(|\mathbf Q|)$};

\end{tikzpicture}
\color{black}
\caption{Three contributing classes to the cancellation of polygon $\mathbf Q$. The \color{orange}left \color{black} and \color{brown}middle \color{black} classes only have support on even $i$. The class on the \color{violet}right \color{black} gives the \color{red}red \color{black} summand for even $i$ and the \color{blue}blue \color{black} summand for odd $i$.}
\label{fig:Even_r_octagon_and_above}
\end{figure}

For $|\mathbf Q|=6$, the contributing terms for a hexagon $\mathbf Q$ are
\begin{equation}
\begin{aligned}
    (\star_j T^A_j)\otimes [q_{\mathbf Q},1] \otimes \bigg[&\color{red}\log(q_{\mathbf Q(1)\,\mathbf Q(5)\,\mathbf Q(2)\,\mathbf Q(6)})+\log(q_{\mathbf Q(4)\,\mathbf Q(3)\,\mathbf Q(5)\,\mathbf Q(2)})\\
    &\color{brown}-\log(q_{\mathbf Q(1)\,\mathbf Q(3)\,\mathbf Q(2)\,\mathbf Q(6)})-\log(q_{\mathbf Q(4)\,\mathbf Q(3)\,\mathbf Q(5)\,\mathbf Q(6)})\\
    &\color{blue}-\log(q_{\mathbf Q(1)\,\mathbf Q(5)\,\mathbf Q(4)\,\mathbf Q(6)})-\log(q_{\mathbf Q(2)\,\mathbf Q(1)\,\mathbf Q(3)\,\mathbf Q(4)})\color{black}+\log q_{\mathbf Q}\bigg]=0\,.
\end{aligned}
\end{equation}
The first six terms arise from the left and right deletions in \eqref{eq:append} for the three quadrangulations of hexagon $\mathbf Q$ illustrated in Figure \ref{fig:Odd_r_hexagon}. The last term arises from the weight-lowering contribution.
This concludes the proof that the operator $K$ annihilates the quasi-shuffle product of all branches in the arborification.

\subsection{Annihilation of Central Polygons}\label{sec:Annihilation_central_polygon}

From \eqref{eq:BFC} and \eqref{eq:append} and the conclusion of Appendix \ref{sec:Annihilation_branches}, we obtain the form
\begin{equation}
\begin{aligned}
    K(T_{\mathbf P}^B) = \sum_{\substack{\mathbf S'\subseteq \mathbf P\\|\mathbf S'|\ge 4}} K(F_{\mathbf S'}\otimes [\crB(\mathbf S'),|\mathbf S'|/4]) = \sum_{\substack{\mathbf S\subseteq \mathbf P\\|\mathbf S|\geq 8}} F_{\mathbf S}\otimes [\crB(\mathbf S),|\mathbf S|/4{-}1]\otimes [\text{final entry}]\,,
\end{aligned}
\end{equation}
where $\mathbf S$ and $\mathbf S'$ are central polygons in $\mathbf P_{4n}$. We drop the overall alternating sign factor $(-1)^n$ in this section for notational simplicity. As in the previous section, we collect all contributions to a particular central polygon $\mathbf S$ and show that the final entry cancels. Following the arborification algorithm, we choose a representative $\mathbf S=(\mathbf S(1),\dots,\mathbf S(|\mathbf S|))$ such that $\mathbf S(1)$ is odd.

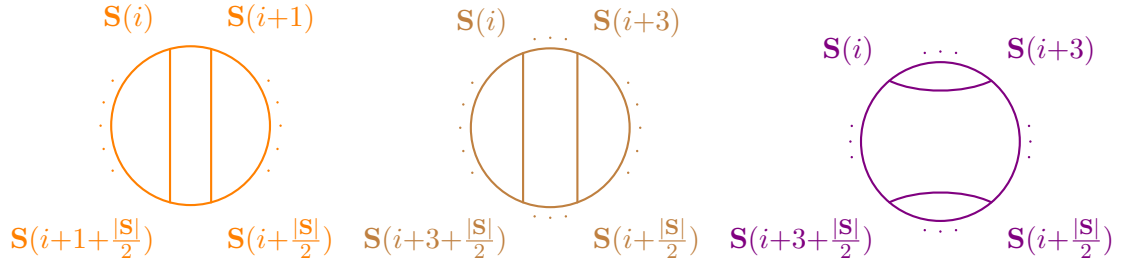
\begin{figure}
\centering
\color{orange}
\begin{tikzpicture}[scale=0.35, line cap=round, line join=round]

% radii
\def\R{3}
\def\Rout{3.4}   % radius for dots outside the circle
\def\rDot{0.045} % size of dots

% circle
\draw[thick] (0,0) circle (\R);

% four boundary points
\coordinate (LT) at (105:\R);  % left top
\coordinate (RT) at ( 75:\R);  % right top
\coordinate (LB) at (255:\R);  % left bottom
\coordinate (RB) at (285:\R);  % right bottom

% straight chords connecting the four boundary points

\draw[thick] (LT) -- (LB);
\draw[thick] (RT) -- (RB);

% dots outside the boundary

% right arc
\foreach \ang in {30,15,0,-15,-30}
  \fill (\ang:\Rout) circle (\rDot);

% left arc
\foreach \ang in {210,195,180,165,150}
  \fill (\ang:\Rout) circle (\rDot);

% labels
\node[above left=2pt]  at (LT) {$\mathbf S(i)$};
\node[below left=2pt]  at (LB) {$\mathbf S(i{+}1{+}\frac{|\mathbf S|}{2})$};
\node[above right=2pt] at (RT) {$\mathbf S(i{+}1)$};
\node[below right=2pt] at (RB) {$\mathbf S(i{+}\frac{|\mathbf S|}{2})$};

\end{tikzpicture}
\color{brown}
\begin{tikzpicture}[scale=0.35, line cap=round, line join=round]

% radii
\def\R{3}
\def\Rout{3.4}   % radius for dots outside the circle
\def\rDot{0.045} % size of dots

% circle
\draw[thick] (0,0) circle (\R);

% four boundary points
\coordinate (LT) at (110:\R);  % left top
\coordinate (RT) at ( 70:\R);  % right top
\coordinate (LB) at (250:\R);  % left bottom
\coordinate (RB) at (290:\R);  % right bottom

% straight chords connecting the four boundary points

\draw[thick] (LT) -- (LB);
\draw[thick] (RT) -- (RB);

% dots outside the boundary

% right arc
\foreach \ang in {30,15,0,-15,-30}
  \fill (\ang:\Rout) circle (\rDot);

% left arc
\foreach \ang in {210,195,180,165,150}
  \fill (\ang:\Rout) circle (\rDot);

% top arc
\foreach \ang in {100,90,80}
  \fill (\ang:\Rout) circle (\rDot);

% bottom arc
\foreach \ang in {280,270,260}
\fill (\ang:\Rout) circle (\rDot);

% labels
\node[above left=2pt]  at (LT) {$\mathbf S(i)$};
\node[below left=2pt]  at (LB) {$\mathbf S(i{+}3{+}\frac{|\mathbf S|}{2})$};
\node[above right=2pt] at (RT) {$\mathbf S(i{+}3)$};
\node[below right=2pt] at (RB) {$\mathbf S(i{+}\frac{|\mathbf S|}{2})$};

\end{tikzpicture}
\color{violet}
\begin{tikzpicture}[scale=0.35, line cap=round, line join=round]

% radii
\def\R{3}
\def\Rout{3.4}   % radius for dots outside the circle
\def\rDot{0.045} % size of dots

% circle
\draw[thick] (0,0) circle (\R);

% four symmetric points on the circle
\coordinate (LT) at (130:\R);  % left top
\coordinate (RT) at ( 50:\R);  % right top
\coordinate (LB) at (230:\R);  % left bottom
\coordinate (RB) at (310:\R);  % right bottom

% top chord
\draw[thick]
(LT) .. controls (-0.9,1.8) and (0.9,1.8) .. (RT);

% bottom chord: symmetric to the top chord
\draw[thick]
(LB) .. controls (-0.9,-1.8) and (0.9,-1.8) .. (RB);

% dots outside the boundary
% top arc
\foreach \ang in {100,90,80}
  \fill (\ang:\Rout) circle (\rDot);

% right arc
\foreach \ang in {10,0,-10}
  \fill (\ang:\Rout) circle (\rDot);

% left arc
\foreach \ang in {190,180,170}
  \fill (\ang:\Rout) circle (\rDot);

% bottom arc
\foreach \ang in {280,270,260}
\fill (\ang:\Rout) circle (\rDot);

% labels
\node[above left=2pt]  at (LT) {$\mathbf S(i)$};
\node[below left=2pt]  at (LB) {$\mathbf S(i{+}3{+}\frac{|\mathbf S|}{2})$};
\node[above right=2pt] at (RT) {$\mathbf S(i{+}3)$};
\node[below right=2pt] at (RB) {$\mathbf S(i{+}\frac{|\mathbf S|}{2})$};

\end{tikzpicture}
\color{black}
\caption{Three contributing classes to the cancellation of polygon $\mathbf S$. The \color{orange}left \color{black} and \color{brown}middle \color{black} classes have support only for odd $i$. The \color{violet}right \color{black} class gives the \color{red}red \color{black} summand for odd $i$ and the \color{blue}blue \color{black} summand for even $i$.}
\label{fig:central_dodecagon_and_above}
\end{figure}

The contributing terms for central polygon $\mathbf S$ are
\begin{equation}
\begin{aligned}
    &F_{\mathbf S}\otimes [\crB(\mathbf S),|\mathbf S|/4-1] \otimes \bigg[\log(\crB(\mathbf S))\color{orange}+\sum_{j=1}^{\frac{|\mathbf S|}{4}}\log \Delta_{\mathbf S(2j-1)\,\mathbf S(2j)}\color{brown}-\sum_{j=1}^{\frac{|\mathbf S|}{4}}\log \Delta_{\mathbf S(2j-1)\,\mathbf S(2j+2)}\\
    &\color{red}+\sum_{j=1}^{\frac{|\mathbf S|}{4}}\log (q_{\mathbf S(2j+1)\,\mathbf S(2j)\,\mathbf S(2j+2)\,\mathbf S(2j-1)})\color{blue}-\sum_{j=1}^{\frac{|\mathbf S|}{4}}\log (q_{\mathbf S(2j)\,\mathbf S(2j+1)\,\mathbf S(2j+3)\,\mathbf S(2j+2)})\color{black}\bigg]=0\,.
\end{aligned}
\end{equation}
The first term is the weight-lowering contribution in \eqref{eq:append}, the first two sums are from left deletion, and the last two sums are from right deletion. The corresponding contributions are illustrated in Figure \ref{fig:central_dodecagon_and_above}. To prove this identity, we use the defining property of the central polygon $\mathbf S(i)=\overline{\mathbf S(|\mathbf S|/2+i)}$.

Under the $\li$ map~(\ref{eq:Li0R}), this concludes the proof that $K \qli_n^B=0$.

\bibliographystyle{JHEP}
\bibliography{biblio.bib}

\end{document}